\documentclass{mn2e}
\usepackage{epsfig}
\usepackage{color}
\usepackage{amsmath}
\title{Methanol and excited OH masers towards W51: I - Main and South}

\author[Etoka, Gray \& Fuller]
{S. Etoka$^1$\thanks{E-mail: Sandra.Etoka@googlemail.com}, 
M.D. Gray$^1$ \& G.A. Fuller$^1$ \\
$^1$Jodrell Bank Centre for Astrophysics, School of Physics and Astronomy, 
The University of Manchester, Manchester M13 9PL, UK \\
}

\begin{document}
\maketitle

%----------------------------------------------------------------------
\begin{abstract}
MERLIN phase-referenced polarimetric observations towards the W51 complex 
were carried out in March 2006 in the Class~II methanol maser transition 
at 6.668~GHz and three of the four excited OH maser hyperfine transitions at 
6~GHz. 
Methanol maser emission is found towards both W51~Main and South.
We did not detect any emission in the excited OH maser lines at 6.030 and 
6.049~GHz down to a 3$\sigma$ limit of $\sim$20~mJy~beam$^{-1}$.
Excited OH maser emission at 6.035-GHz is only found towards W51~Main.
This emission is highly circularly polarised 
(typically $\ge 45$~\% and up to 87\%). Seven Zeeman pairs were 
identified in this transition, one of which contains detectable linear 
polarisation. The magnetic field strength derived from these Zeeman pairs 
ranges from $+1.6$ to $+6.8$~mG, consistent with the previously published 
magnetic field strengths inferred from the OH ground-state lines. 
The bulk of the methanol maser emission is associated with 
W51~Main, sampling a total area of {$\sim 3\arcsec \times 2.2$\arcsec}
(i.e., $\sim16200 \times 11900$~AU), while only two maser components, 
separated by {$\sim$2.5\arcsec}, are found in the W51~South region. 
The astrometric distributions of both 6.668-GHz methanol and 6.035-GHz 
excited-OH maser emission in the W51~Main/South region are presented here. 
The methanol masers in W51~Main show a clear coherent velocity and spatial 
structure with the bulk of the maser components distributed 
into 2 regions showing a similar conical opening angle with 
of a central velocity of $\sim$+55.5~km~s$^{-1}$ and an expansion velocity of 
$\le$5~km~s$^{-1}$. The mass contained in this structure is estimated to be at 
least 22~M$_\odot$.
The location of maser emission in the two afore-mentioned lines is 
compared with that of previously published OH ground-state emission. 
Association with the UCHII regions in the W51~Main/South complex 
and relationship of the masers to infall or outflow in the region are 
discussed.
\end{abstract}


%----------------------------------------------------------------------
\section{Introduction}

W51, originally detected by Westerhout (1958), is one of the most massive 
Giant Molecular Clouds (GMCs) in the Galaxy (Mufson \& Liszt 1979).
This $1.6 \times 10^6$~M$_{\odot}$ GMC (Carpenter \& Sanders 1998) is
a very rich high-mass star-forming region (SFR) environment, located in the 
Sagittarius spiral arm, which contains two giant HII regions labelled W51A and 
W51B which are themselves resolved into smaller components (Mehringer 1994).
Within W51A, the most luminous site labelled W51e, where W51~Main/South 
is found, is one of the brightest regions.  
Trigonometric parallax measurements of 
the 22~GHz H$_2$O maser source in W51 Main/South
from Sato et al. (2010) 
give a distance of $5.4 \pm 0.3$~kpc for the complex. 
W51~Main/South has been studied at a wide range of wavelengths both in 
continuum and molecular (thermal and maser) transitions.
It is known to host many ultra-compact (UC) 
HII regions thought to contain massive young stellar objects  at various 
stages of evolution: the UCHII regions are labelled W51e1, e2, e3 and e4 
(Gaume, Johnston \& Wilson 1993). Of these, W51e2 was found to be composed of 
four more compact continuum sources labelled W51e2-W, e2-E, e2-NW and e2-N 
(Shi, Zhao \& Han 2010a), and the UCHII region labelled W51e8 
(Zhang \& Ho 1997).
Zhang, Ho, \& Ohashi (1998) 
found evidence of gravitational infall towards W51e2 and W51e8. 
Kalenskii \& Johansson (2010) performed a survey at 3~mm towards W51e1/e2
covering the frequency range 84-115~GHz that detected 105 molecules and 
their isotopic species.
Maser emission from H$_2$O, ground-state OH, NH$_3$ and 
methanol has also been found
(Genzel et al. 1978 \& 1981; Menten, Melnick \& Phillips 1990a; 
Menten et al. 1990b; Imai et al. 2002; 
Fish \& Reid 2007; Zhang \& Ho 1997; Phillips \& van~Langevelde 2005). 
Excited OH maser emission at 6.035~GHz towards W51 has also been detected 
and observed on several occasions (Rickard, Zuckerman \& Palmer 1975; 
Caswell \& Vaile 1995; Desmurs \& Baudry 1998) but its actual provenance in 
the SFR complex has not been ascertained yet. \\

We have performed MERLIN astrometric observations of excited-OH maser emission
at 6.035, 6.030 and 6.049~GHz and Class~II methanol maser emission at 
6.668~GHz towards W51 to investigate the relationship between the maser 
emission and the compact continuum sources in this SFR complex.
Here we present the results regarding W51~Main/South. 
The other regions of the W51 complex will be presented in a subsequent paper.
We compare the distribution of the detected 6.035-GHz excited-OH and 6.668-GHz 
methanol maser emission to that of the ground-state OH in the same 
region. Additional information regarding the magnetic field strength 
and structure of the region is also discussed. The details of the observations 
and the data reduction are presented in
section~\ref{sec: observations and data reduction}. The results 
are presented in section~\ref{sec: results}.
The OH maser modelling carried out for the present study is 
presented in section~\ref{sec: modelling}. A discussion of the results is given
in section~\ref{sec: discussion}. A summary and conclusions are presented 
in section~\ref{sec: conclusion}.

%----------------------------------------------------------------------
\section{Observations and data reduction}
\label{sec: observations and data reduction}

Observations were performed using the six telescopes of MERLIN available at 
5~cm at that time (namely Defford, Cambridge, Knockin, Darnhall, 
MK2 \& Tabley) in full polarimetric mode.
The two excited OH maser main lines at a rest frequency of
6.030747 and  6.035092~GHz 
were observed on 2006 March 21 while 
the satellite OH maser line at a rest frequency of 6.049084~GHz
and the methanol maser transition at 
a rest frequency of 6.668518~GHz 
were observed on 2006 March 20, 
25 and 30. For all the transitions, a spectral bandwidth of 0.5~MHz divided 
into 512 channels was used, giving a velocity coverage of 
22.745 and 24.855~km~s$^{-1}$
and a channel spacing of 0.044 and 0.049~km~s$^{-1}$
for the 6.668~GHz methanol and 6.035~GHz excited OH maser datasets, 
respectively.
The central velocity was taken to  be 
$v_{\rm LSR}=+55.50$~km~s$^{-1}$ 
for all observations. 
All the velocities given in this paper are relative to the local
standard of rest (LSR).
The observations of the target at all transitions were interspersed with short 
periods on the same phase reference calibrator, 1920+154. 
3C84 was used as the bandpass calibrator and for the 
polarisation leakage (d-term) corrections.
3C286 was used to retrieve the absolute polarisation
position angles and to provide the flux density reference. 
The data reduction followed the procedure explained in section~2 of 
Etoka, Cohen \& Gray (2005). \\

The initial data editing, the gain-elevation correction and a
first-order amplitude calibration were applied using the {\tt d-program},
a MERLIN-specific package available at Jodrell (Diamond et al. 2003). 
Further refined data editing, the remaining instrumental 
and atmospheric effect calibrations, bandpass and second order amplitude 
calibrations and phase referencing were performed within the {\sc aips} package 
(Diamond et al. 2003, Appendix~F). \\

The final phase calibration on the source was performed using the brightest 
channel in the RHC polarisation of the 6.035~GHz excited OH dataset with 
transfer of solutions onto the LHC polarisation. For the 6.668~GHz methanol 
dataset the second strongest channel was used as it led to a better 
phase calibration. With an intensity ratio of roughly 10 between the 
reference channel adopted and the brightest channel we estimate that the 
intensity inferred for the strongest component is underestimated by 
$\sim$10\%. \\

The accuracy in absolute position for the masers
is limited by four factors: (1) the position
accuracy of the phase calibrator, (2) the accuracy of the telescope
positions, (3) the relative position error proportional to the
beamsize and inversely proportional to the signal-to-noise ratio, and
finally (4) the atmospheric variability. The first two factors are
frequency independent and were estimated to be $\sim$5~mas (given by the
MERLIN calibrator catalogue) and 10~mas (Diamond et al. 2003) respectively. 
The third factor is very small in comparison to the other sources of
error (2-3~mas in the worst case). The fourth error is
estimated from the quality of the phase of the calibrator and its
separation from the source.  
Given the angular separation of {1.2\degr}, we estimate that this final error 
is less than 8~mas. All these errors add quadratically to give
14~mas as the total systematic error in the {\em absolute} positions measured
in this paper. \\

All the offset positions given from here onwards are with respect to the 
reference position, taken to be that of Comp~24 of the methanol 6.668~GHz 
dataset (marked with an asterisk in Table~\ref{Table: 6.668 GHz met info}), 
namely, 
          RA$_{\rm J2000}$=$19^{\rm h}23^{\rm m}43\fs948$ 
          Dec$_{\rm J2000}$=14\degr30\arcmin34\farcs38 $\pm 14$~milliarcsec. 
Given that the 6-GHz excited OH and 6.668-GHz methanol 
datasets were obtained with the same instrument a few days appart, 
an extra quadratic atmospheric variability term needs to be taken into 
consideration for the calculation of the {\em alignment} uncertainty.
Consequently, the accuracy of the {\em alignment} of the two maser types is 
$16$~mas. \\

The source was imaged
with the task {\sc imagr} using a pixel separation of 
5~mas. The restoring beamwidth at 6.668~GHz was 
78~$\times$~48~mas$^{2}$ with a position angle (P.A.) of  22$\degr$.
At 6.035~GHz the restoring beamwidth was 64~$\times$~45~mas$^{2}$ 
with a P.A. of 23$\degr$.
Typically, the rms noise level ($\sigma$), obtained from the image of a channel 
free of emission, was about 7~mJy~beam$^{-1}$ for the 6.668-GHz methanol 
dataset, increasing to 150~mJy~beam$^{-1}$ in the channel with the strongest 
emission. 
For the 6.035~GHz excited OH dataset the $\sigma$ was typically 14~mJy 
increasing to 18~mJy in the channel with the strongest emission.\\

The datacubes were inspected and the maser emission features were 
fitted with two-dimensional Gaussians  
using the task {\sc sad} with a flux threshold of 3$\sigma$.  
Maser spots were grouped as `components' for further analysis if they 
showed emission in at least three consecutive frequency channels with 
relative positional offsets of less than 25~mas.   \\

Polarisation information was retrieved for the 6-GHz excited OH maser 
emission. At the position of each maser component found in the Stokes~$I$ 
data cube, the flux in Stokes~$Q$, $U$ and $V$ were measured. 
The linear polarised intensity $P = \sqrt{Q^2+U^2}$ and 
polarisation angle $\chi = 0.5 \times$arctan($U/Q$) were then calculated for 
each maser component. $P$ is considered significant only if it is greater 
than 3$\sigma$. \\

%----------------------------------------------------------------------
\section{Results}
\label{sec: results}

%-  -  -  -  -  -  -  -  -  -  -  -  -  -  -  -  -  -  -  -  -  -  -  -
\subsection{Methanol maser emission at 6.668~GHz} 
\label{subsec: methanol at 6.668 GHz}

Table~\ref{Table: 6.668 GHz met info} presents in decreasing velocity order 
the components detected in the 6.668-GHz methanol maser transition over the 
W51 Main/South region. Column~1 gives the component number, Column~2 gives its 
intensity-weighted velocity, Column~3 gives its specific intensity and 
Columns~4 and 5 give its RA and Dec offset from the reference position. \\

%- - - - - - - - - - - - - - - - - - - - - - - - - - - - - - - - - - -
\begin{table}
 \caption {\small 6.668-GHz methanol maser components}
 \label{Table: 6.668 GHz met info}  
{\footnotesize
 \begin{tabular}{rrrrr}
 \hline
 Comp & Vel & I & $\Delta \alpha$$^a$ & $\Delta \delta$$^a$  \\
{\tiny   Nb    } & {\tiny (km~s$^{-1}$)}  & {\tiny (Jy~b$^{-1}$)} &
{\tiny (mas)} & {\tiny (mas)}  \\
 \hline   
 1  &  64.046 &   0.285  &  -1558.7  & -9590.8 \\
 2  &  60.841 &   0.182  &    862.4  &   177.7 \\
 3  &  60.434 &   2.853  &    206.8  &   177.8 \\
 4  &  60.151 &   8.622  &     11.9  &    43.8 \\
 5  &  60.142 &   0.449  &    866.0  &   501.5 \\
 6  &  60.019 &  24.462  &      5.8  &    57.8 \\
 7  &  59.290 & 216.410  &      3.5  &    66.6 \\
 8  &  58.821 &  26.625  &     67.0  &   -79.0 \\
 9  &  58.807 &   2.654  &     16.9  &   171.7 \\
10  &  58.796 &   0.986  &    132.4  &  -158.1 \\
11  &  58.738 &   1.347  &     72.4  &   249.1 \\
12  &  58.736 &   1.337  &    274.3  &    88.5 \\
13  &  58.712 &   2.936  &      9.8  &   -18.6 \\
14  &  58.684 &   2.002  &      3.3  &   144.3 \\
15  &  58.672 &   1.814  &     37.4  &    35.7 \\
16  &  58.555 &   0.809  &     69.3  &    66.8 \\
17  &  58.515 &   1.458  &    925.3  &   162.9 \\
18  &  58.413 &   0.550  &    279.0  &   -91.2 \\
19  &  58.393 &   1.377  &    157.5  &    52.2 \\
20  &  58.390 &   0.656  &    -29.1  &   126.1 \\
21  &  58.317 &   3.177  &    681.0  &   414.2 \\
22  &  58.238 &   4.221  &    274.5  &   100.3 \\
23  &  58.231 &   0.959  &    273.2  &   110.2 \\
$^*$24  &  58.007 &  28.387  &      0.0  &     0.0 \\
25  &  57.429 &   0.298  &    942.3  &   153.9 \\
26  &  57.326 &   0.900  &   -356.9  &  2273.1 \\
27  &  56.805 &   0.117  &   1012.8  &  1077.0 \\
28  &  56.786 &   0.092  &    976.9  &  1093.6 \\
29  &  55.331 &   0.360  &   -895.5  & -7103.6 \\
30  &  54.954 &   0.230  &   -484.5  &  -148.0 \\
31  &  53.008 &   0.237  &   -737.1  &   378.4 \\
32  &  52.888 &   0.115  &   -298.0  &  -289.1 \\
33  &  52.811 &   0.198  &  -1024.9  &    22.2 \\
34  &  52.778 &   0.549  &    717.3  &  1354.4 \\
35  &  52.593 &   0.331  &  -1003.3  &    42.7 \\
36  &  52.476 &   0.718  &  -1025.8  &    31.1 \\
37  &  52.371 &   1.611  &   -366.9  &  -434.4 \\
38  &  52.248 &   0.629  &  -1003.4  &   244.4 \\
39  &  52.175 &   0.146  &   -947.4  &    37.6 \\
40  &  52.157 &   0.141  &   -181.4  &  -381.5 \\
41  &  51.947 &   4.288  &   -193.6  &  -239.8 \\
42  &  51.880 &   0.994  &   -387.7  &  -473.5 \\
43  &  51.860 &   0.728  &   -981.0  &   248.3 \\
44  &  51.631 &   1.077  &    621.9  &  1030.8 \\
45  &  51.525 &   2.174  &   -392.0  &  -467.9 \\
46  &  51.500 &   0.308  &   -787.0  &  -291.9 \\
47  &  51.053 &   0.599  &   -199.8  &  -239.0 \\
48  &  51.018 &   0.150  &   -900.6  &   223.1 \\
49  &  50.904 &   0.247  &   -422.1  &  -480.0 \\
50  &  49.934 &   0.198  &    623.3  &   969.2 \\
 \hline
 \end{tabular} \\
}
{\flushleft
\noindent $\bf a:$ $\Delta \alpha$ and $\Delta \delta$ are relative to  
          the reference position taken to be that of Comp~24 (marked with an 
          asterisk):

          RA$_{\rm J2000}$=$19^{\rm h}23^{\rm m}43\fs948$ 
          Dec$_{\rm J2000}$=14\degr30\arcmin34\farcs38.
}
\end{table}
%- - - - - - - - - - - - - - - - - - - - - - - - - - - - - - - - - - -

\begin{figure*}
\hspace*{-1cm} \epsfig{file=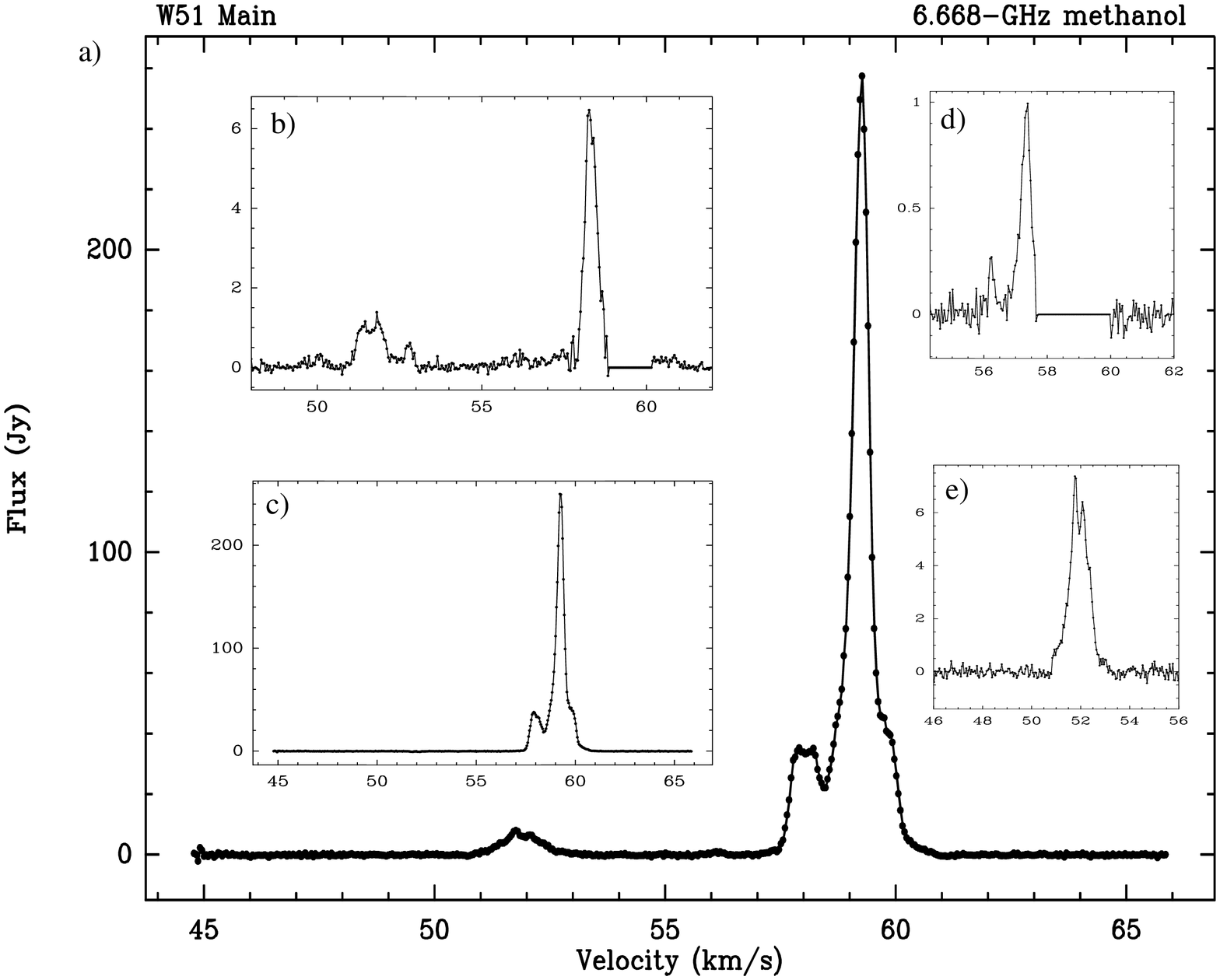,angle=-90,width=19.5cm}
%%\vspace{-2.5cm}
 \caption{Spectra constructed from the final 
          images
          for the various regions 
          where 6.668-GHz methanol maser emission has been detected 
          towards W51~Main.
          {\bf a)} Spectrum for the entire region covering an area of 
          $\sim${3\arcsec}~$\times$~{2.2\arcsec}.
          {\bf b)} Spectrum for the north-east region covering the area
          $\Delta \alpha$=[+1000,+500]~mas $\Delta \delta$=[+100,+1400]~mas.
          {\bf c)} Spectrum for the central region covering the area
          $\Delta \alpha$=[+500,-100]~mas $\Delta \delta$=[-200,+300]~mas.
          {\bf d)} Spectrum for the northern region centered around
          $\Delta \alpha=-360$~mas $\Delta \delta=+2270$~mas.
          {\bf e)} Spectrum for the south-west region covering the area
          $\Delta \alpha$=[-100,-1100]~mas $\Delta \delta$=[-500,+400]~mas.
          Note that the channels corresponding to the strongest peak 
          covering the velocity range $\sim$[+59.0,+60.0] and 
          $\sim$[+57.7,+60.5]~km~s$^{-1}$ have been 
          set to zero for the spectra in $\bf b$ and $\bf d$
          respectively due to the high level of noise in the corresponding 
          images.
 }
 \label{fig: Methanol ISPEC W51Main}
 \end{figure*}

\begin{figure*}
\hspace*{-1cm} \epsfig{file=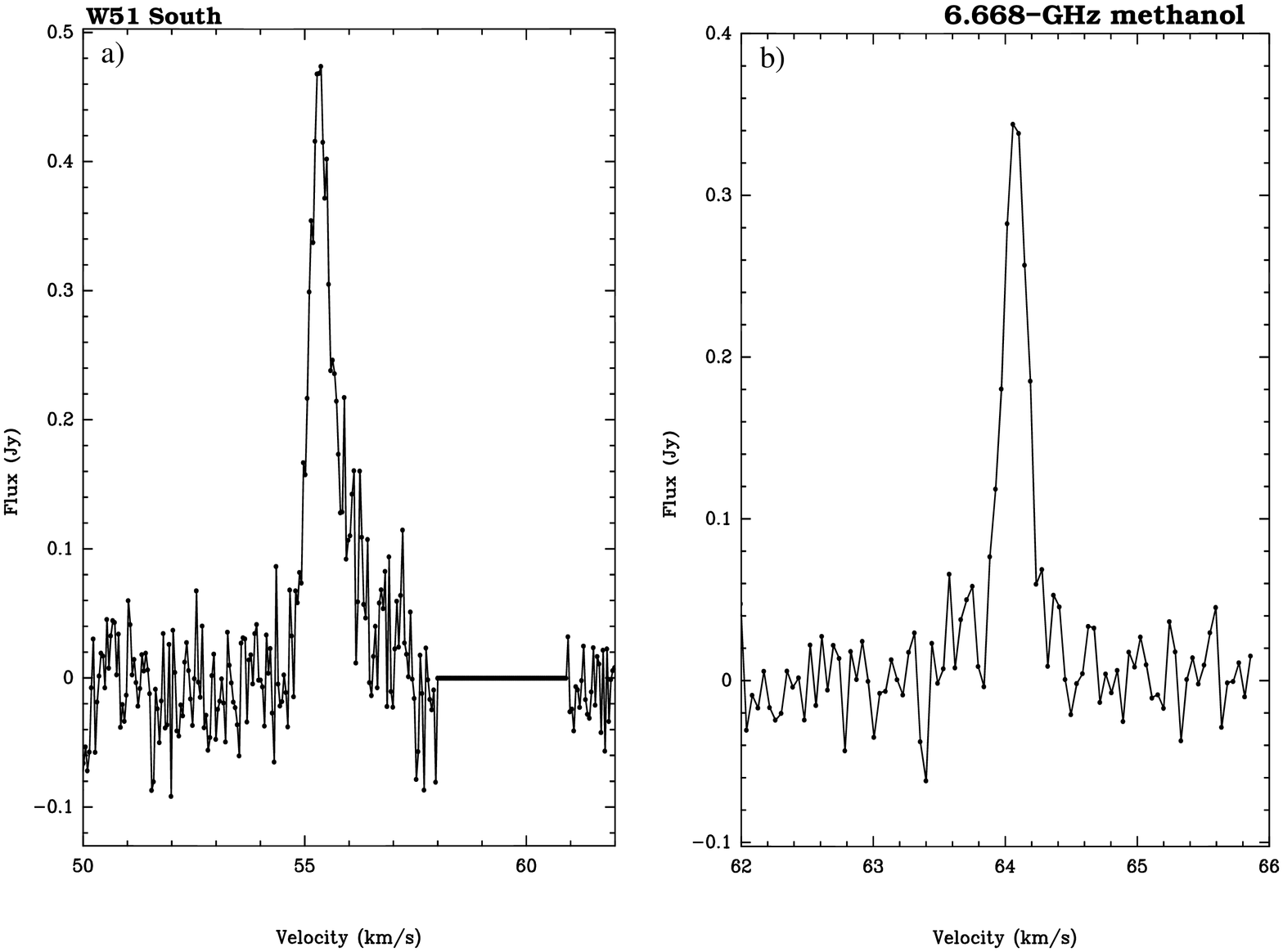,angle=-90,width=19.5cm}
%%\vspace{-2.5cm}
 \caption{Spectra constructed from the final 
         images
         for the two regions 
         where 6.668-GHz methanol maser emission has been detected 
         towards W51~South, covering in both cases an area of 
         $\sim 250 \times 250$~mas$^2$. 
         {\bf a)} Spectrum for the northern part. Note that the channels 
             corresponding to the strongest peak covering the velocity range 
             $\sim$[+58,+60.5]~km~s$^{-1}$ have been set to zero 
             due to the high level of noise in the corresponding 
             images.
         {\bf b)} Spectrum for the southern part.}
 \label{fig: Methanol ISPEC W51South}
 \end{figure*}

Figures~\ref{fig: Methanol ISPEC W51Main} 
and~\ref{fig: Methanol ISPEC W51South} 
present the spectra constructed from the final 
images for the regions, 
seemingly distinct in the plane of the sky, where 6.668~GHz methanol maser 
emission has been detected towards W51~Main and W51~South respectively. 
Unlike W3(OH) where a large-scale methanol maser filament 
has been found (Harvey-Smith \& Cohen 2006) such an extended emission is not 
detected.
Comparison of the spectral profile displayed in the main panel of 
Figures~\ref{fig: Methanol ISPEC W51Main} with that obtained    
with the Australia Telescope Compact Array (ATCA) in May 2007 
(Avison, PhD 2010), with an angular resolution of $\sim 40\arcsec$, shows 
a good agreement regarding the main spectral group of components observed 
and their relative intensity ratio, indicating that there is no substantial 
resolved emission 
Towards W51~Main, the overall area from which 6.668-GHz methanol maser is 
emitted covers 
$\sim${3\arcsec}~$\times$~{2.2\arcsec} (i.e., $\sim16200 \times 11900$~AU),
and can be divided roughly into four areas. 

In W51~Main there seem to be 4 distinct regions of emission on the plane of 
the sky.
The bulk of the emission arises from the multi-component peak in the velocity 
range 
[+57.0,+61]~km~s$^{-1}$ 
and comes from a region in W51~Main 
covering the area $\Delta \alpha$=[+500,-100]~mas 
$\Delta \delta$=[-200,+300]~mas (hereafter, the ``central region''; 
Fig.~\ref{fig: Methanol ISPEC W51Main}c). The spectrum of the region 
located north-east of the central region and covering the area 
$\Delta \alpha$=[+1000,+500]~mas $\Delta \delta$=[+100,+1400]~mas 
(Fig.~\ref{fig: Methanol ISPEC W51Main}b) shows the largest spread in 
velocity, with a total velocity range of $\sim8$~km~s$^{-1}$. It is made of
two main groups of components centred at 
$\sim +52.0$~km~s$^{-1}$ and $\sim +58.2$~km~s$^{-1}$.
The latter group peak velocity is encompassed by
the velocity range covered by the bluest group of components of the central 
region. The spectrum of the south-west region covering the 
area $\Delta \alpha$=[-100,-1100]~mas $\Delta \delta$=[-500,+400]~mas 
(Fig.~\ref{fig: Methanol ISPEC W51Main}e) encompasses a similar 
velocity range as that of the north-east blue peak. The spectrum from the 
faint group of components emanating from the northern region centered 
around $\Delta \alpha=-360$~mas $\Delta \delta=+2270$~mas 
(Fig.~\ref{fig: Methanol ISPEC W51Main}d) covers the bluer edge of the 
prominent multi-component peak.

Towards W51~South, only two localised areas are found to be producing 
6.668-GHz methanol maser emission. \\

%%%%%%%%%%%%
%%%%%%%%%%%%
%%%%%%%%%%%%
%%%%%%%%%%%%

%%%%%%%%%%%%
%%%%%%%%%%%%
%%%%%%%%%%%%
%%%%%%%%%%%%

%%%%%%%%%%%%
%%%%%%%%%%%%
 \begin{figure*}
\hspace*{-1cm} \epsfig{file=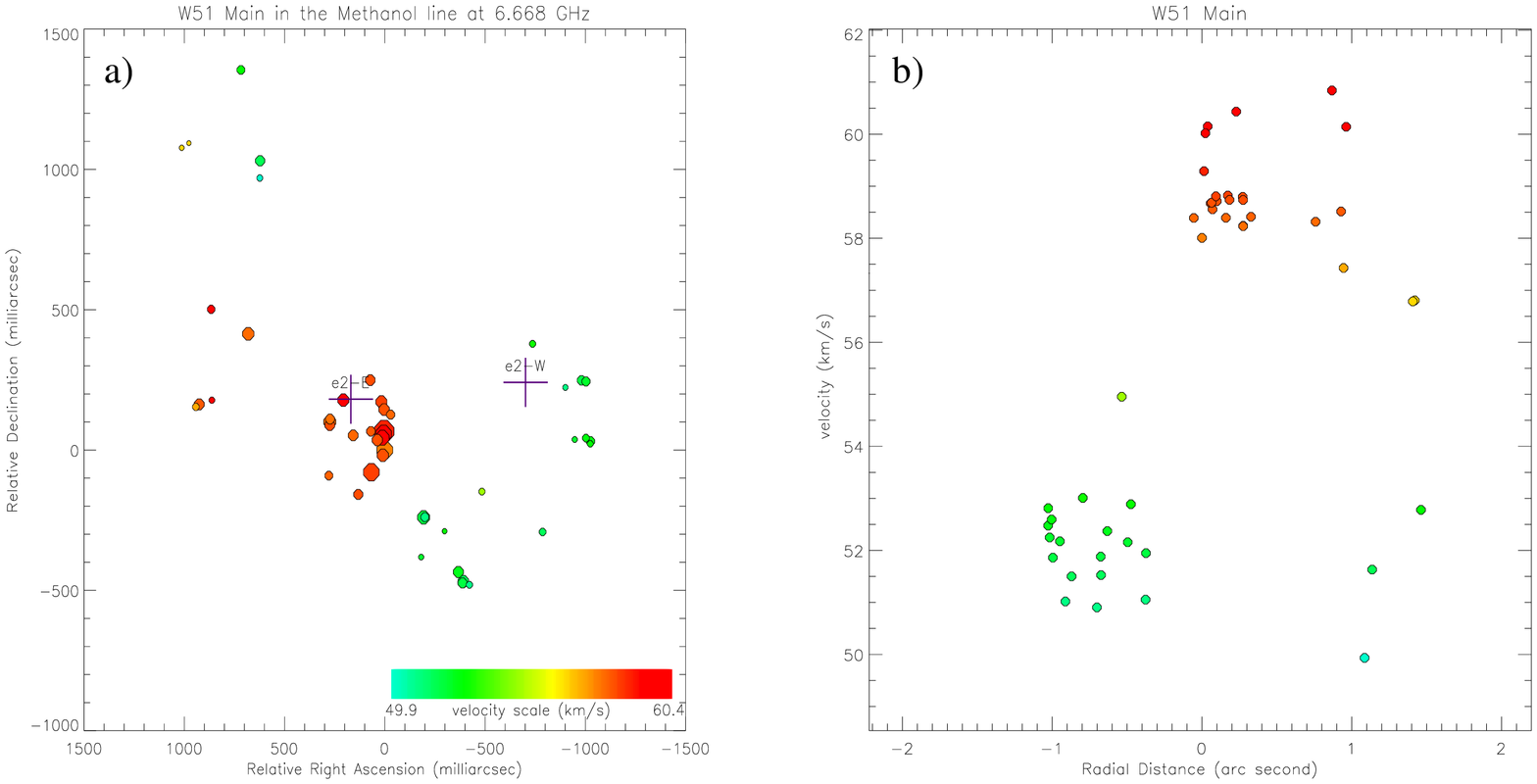,angle=-90,width=19.5cm}
\vspace{-2.5cm}
 \caption{{\bf a)} Maser components in the methanol 6.668-GHz line towards 
          W51~Main, excluding the northern component, Comp~26, at the RA and 
          Dec offsets 
          $\Delta \alpha \sim -357$~mas and $\Delta \delta \sim +2273$~mas.
          The size of the symbols is proportional to the log of the intensity.
          The velocity colour-code is restricted to the W51~Main methanol maser 
          emission velocity range.
          The crosses indicate the positions of the compact continuum sources 
          in the region (Shi et al. 2010a).
          {\bf b)} The velocity distribution of the maser 
          spot versus radial distance. The central position is taken to be 
          [0,+80~mas] inferred from the convergent point of the SW blue-shifted 
          maser components.}
 \label{fig: Methanol components W51Main}
 \end{figure*}
%%%%%%%%%%%%
%%%%%%%%%%%%

Following the criteria given in 
section~\ref{sec: observations and data reduction},
50~maser components were identified towards the W51~Main/South complex, 
only 2 of which are associated with W51~South.
Figure~\ref{fig: Methanol components W51Main}a presents the maser component 
distribution in the methanol 6.668-GHz transition in W51~Main found around 
W51e2-E and W51e2-W. 
The velocity spread is 10.5~km~s$^{-1}$ with a mean velocity at 
+55.5~km~s$^{-1}$. 
The masers are not uniformly distributed but seem to be part of 3 main 
regions, two of which are seemingly distributed into ellipsoidal structures 
reminscent of what is observed in various SFR complexes by Bartkiewicz et al. 
(2009). The overall maser emission is roughly oriented along a 
P.A.~$\simeq 70^{\degr}$ (east of north).
The extent of the two largest ellipsoidal 
structures is similar: {$\sim$1.2\arcsec} (i.e., $\sim$6500~AU) while that 
located at the centre is much smaller: {$\le$0.5\arcsec} (i.e., $\le$2700~AU). 
Figure~\ref{fig: Methanol components W51Main}b 
presents the velocity distribution of the maser components versus the radial 
distance from a central position of [0,+80]~mas 
which has been inferred from the converging points of the SW blue-shifted 
maser components. This position is within 15~mas of the strongest 
methanol maser spot (Comp~7 centered at 
$+59.3$~km~s$^{-1}$ 
cf. Table~\ref{Table: 6.668 GHz met info}), 
and in good agreement with the location of W51e2-E given
the beam size of the SMA at 0.85~mm, 0\farcs3 $\times$ 0\farcs2.
There is a clear position-velocity relation for the bulk of 
the maser components found to be distributed mainly into 2 regions 
showing a similar conical opening angle indicative 
of a central velocity of 
$\sim$+55.5~km~s$^{-1}$
and an expansion velocity of 
$\le$5~km~s$^{-1}$.

It has to be noted that the systemic velocity  
for W51e2 estimated from the different hot molecular lines varies significantly,
from $\sim +54$ to $\sim +59$~km~s$^{-1}$ 
(e.g., Shi et al. 2010a; Sollins, Zhang \& Ho 2004; Remijan et al. 2004; 
Zhang et al. 1998; Rudolph et al. 1990).
The 
$\sim$+55.5~km~s$^{-1}$
convergent central velocity inferred from 
Fig.~\ref{fig: Methanol components W51Main}b is in close agreement with the 
central velocity of the HCN(4-3) absorption line observed towards 
W51e2-E by Shi, Zhao \& Han (2010b, their Fig~4b) and interpreted as being 
redshifted absorption of HCN against the compact continuum core. \\

The five components forming the northern part of the north-east ellipsoidal 
structure (covering the area $\Delta \alpha$=[+1000,+500]~mas 
$\Delta \delta$=[+900,+1400]~mas in the north-east region observed in 
Fig.~\ref{fig: Methanol components W51Main}a and found at the radial 
distance [+1.0{\arcsec},+1.6{\arcsec}] in 
Fig.~\ref{fig: Methanol components W51Main}b) do not follow the 
velocity-position relation. As noted previously, the spectrum 
corresponding to the north-east ellipsoidal structure is the most extended 
one with a total velocity extent of $\sim$8~km~s$^{-1}$ 
(cf. Fig~\ref{fig: Methanol ISPEC W51Main}b). These two facts seem to indicate 
that these masers, though appearing to probe a distinct ellipsoidal structure 
in the plane of the sky, actually trace different physical components in the 
W51~Main region. Moreover, the velocity distribution versus the radial distance 
(Fig.~\ref{fig: Methanol components W51Main}b) indicates that the southern 
redshifted part of this plane-of-the-sky structure (covering the area 
$\Delta \alpha$=[+1000,+500]~mas $\Delta \delta$=[+100,+500]~mas) 
is dynamically associated with the ``central region''. \\

One maser component (Comp~26 in Table~\ref{Table: 6.668 GHz met info}),
centered at a velocity of $+57.3$~km~s$^{-1}$, is found {$\ge$2\arcsec} 
north of the bulk of the maser components in W51~Main, seemingly
associated with either W51e2-N or W51e2-NW (0.5 and {0.65\arcsec} away from 
these cores respectively). \\

\begin{figure*}
\hspace*{-1cm} \epsfig{file=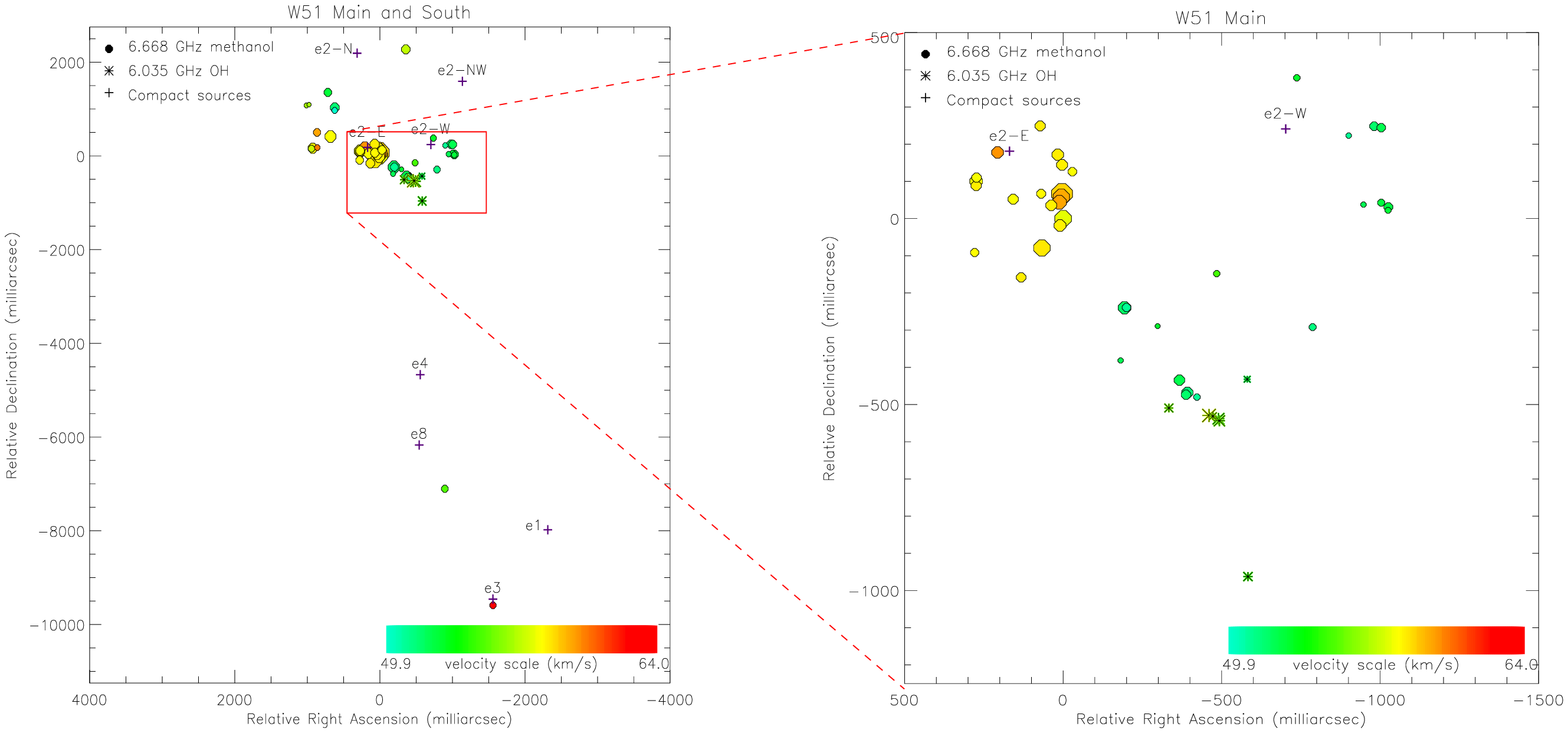,angle=-90,width=19.5cm}
\vspace{-2.5cm}
 \caption{Left panel: Maser components in the methanol 6.668-GHz and the 
          excited OH 6.035-GHz transitions for the overall W51~Main/South 
          region. Right panel: Magnification of part of W51~Main.
          The velocity colour-code covers the W51~Main and South methanol maser 
          emission velocity range.
          The size of the symbols is proportional to the log of the intensity.
          The crosses indicate the positions of the compact continuum sources 
          in the region 
          (Gaume et al. 1993, Zhang \& Ho 1997, 
          Shi, Zhao \& Han 2010a).
}
 \label{fig: Methanol components W51MS and Zoom}
 \end{figure*}

The two maser components detected in W51~South (Comp~1 and Comp~29 in 
Table~\ref{Table: 6.668 GHz met info}) are centered at  
$\sim +64.0$~km~s$^{-1}$ and $\sim +55.3$~km~s$^{-1}$ respectively
(cf. Fig~\ref{fig: Methanol components W51MS and Zoom} and 
Fig~\ref{fig: Methanol ISPEC W51South}). 
The northern component, Comp~29, is closer to W51e8 ($\sim$0.5\arcsec) than 
it is to W51e1 ($\sim$1.1\arcsec) and is consequently more likely to be 
associated with W51e8, though the systemic velocity of this core has 
been estimated to be $\sim +59$~km~s$^{-1}$ by Zhang et al. (1998).
The southern component, Comp~1, is clearly associated with W51e3.

%%%%%%%%%%%%
%%%%%%%%%%%%
%%%%%%%%%%%%
%%%%%%%%%%%%

%-  -  -  -  -  -  -  -  -  -  -  -  -  -  -  -  -  -  -  -  -  -  -  -
\subsection{Excited OH emission} \label{subsec: excited OH}

The theoretical models by Gray, Field \& Doel (1992) and 
Pavlakis \& Kylafis (2000) 
show that the SFR regions are conducive to inversion of the 5-cm main lines 
(overlapping the range of conditions  leading to the inversion of the 
ground-state main lines) and weak inversion of the satellite transition at 
6.049~GHz, but unfavourable for 6.017~GHz inversion. \\

Observational studies show the presence of 6.035-GHz and to a lesser extent 
6.031-GHz maser emission towards a wide range of high-mass SFR sites
known to exhibit ground-state 1.665~GHz maser emission
(e.g., Caswell 2003; Desmurs \& Baudry 1998; Baudry et al. 1997; 
 Caswell \& Vaile 1995). Maser emission from these excited OH maser 
transitions is often present in regions similar to those that emit in the 
Class~II methanol maser at 6.668~GHz (e.g., Green et al. 2007; 
Etoka et al. 2005). 
This is corroborated by the models by Cragg, Sobolev \& Godfrey (2002) which 
show that very similar high-density, low-temperature conditions with a 
substantial radiation field at a higher temperature are needed 
for the inversion of these maser transitions.  \\

We observed the three 6-GHz hyperfine transitions for which inversion in 
SFR regions can potentially occur.
We did not detect any excited OH maser lines at 6.031 or 6.049~GHz towards the 
entire W51 complex down to a 3$\sigma$ limit of $\sim$20~mJy~beam$^{-1}$.  \\

\subsubsection{6.035~GHz} \label{subsec: OH at 6.035 GHz}

\begin{figure}
      \epsfig{file=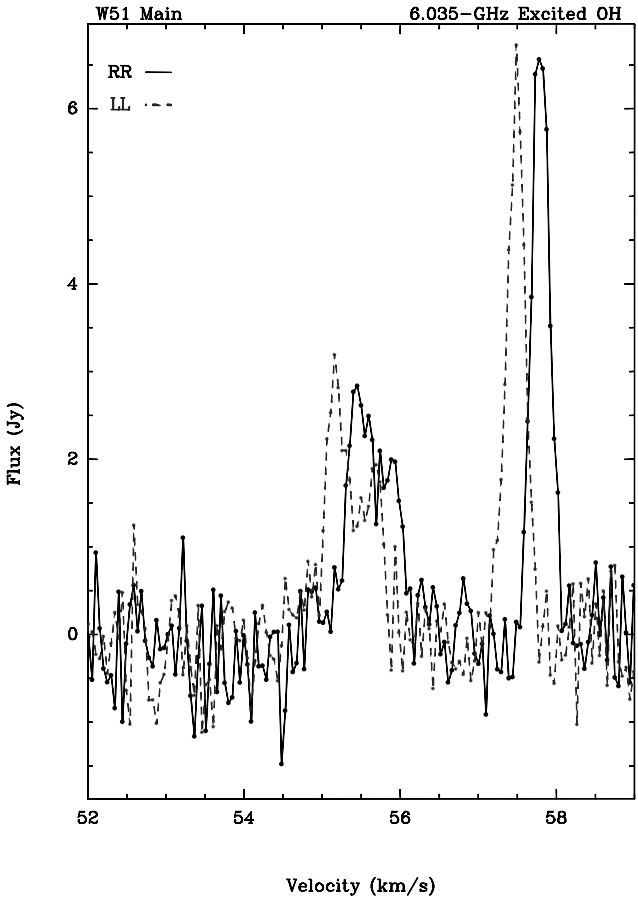,width=9.00cm}
\caption{Spectra in the RHC (solid line) and LHC (dashed line) polarisations 
         constructed from the final images over the entire area where 
         6.035-GHz excited OH emission is found towards W51~Main.}
\label{fig: spectra OH 6035}
\end{figure}

Excited OH emission at 6.035~GHz towards the W51 complex was first detected by 
Rickard et al. (1975) and then reobserved on several occasions  
(Caswell \& Vaile 1995; Desmurs \& Baudry 1998). Two main 
spectral features
are visible in the spectra of Rickard et al. (1975, their Fig.~4) covering a 
velocity range of $+52.8$ to $+56$~km~s$^{-1}$ with the strongest emission 
at 
$\sim +53$~km~s$^{-1}$. 
The spectra obtained by Caswell \& Vaile (1995)
clearly show five 
spectral features
ranging from $+52.8$ to $+64$~km~s$^{-1}$ with the 
strongest 
spectral feature
at $\sim +55$~km~s$^{-1}$.
Figure~\ref{fig: spectra OH 6035} presents the 6.035-GHz excited OH spectra
in the RHC (solid line) and LHC (dashed line) polarisations towards 
W51~Main from the present work. Two 
spectral features at $\sim +53$ 
and 
$\sim +57.5$~km~s$^{-1}$ are clearly visible, with the stronger 
spectral feature at $\sim +57.5$~km~s$^{-1}$ 
reaching a peak flux density of 6~Jy. 
Comparing the spectral profile observed in the 1970's (Rickard et al.), 
1990's (Caswell \& Vaile) and in the 2000's (present work) it is clear that 
the emission is highly variable and is likely the explanation of the 
non-detection of the 2 faint 
spectral features at $\sim +59$
and $\sim +64$~km~s$^{-1}$
detected by Caswell \& Vaile (1995). 
\\

%%%%%%%%%%%%
%%%%%%%%%%%%
%%%%%%%%%%%%
%%%%%%%%%%%%

\begin{table}
 \caption {\small 6.035-GHz RHC and LHC maser components}
 \label{Table: 6.035 GHz RHC and LHC comp info}  
{\footnotesize
 \begin{tabular}{rrrrrl}
 \hline
Comp  & Vel  & Flux & $\Delta \alpha$$^a$ & $\Delta \delta$$^a$ &   Z$^b$ \\
      &  {\tiny (km~s$^{-1}$)}  & {\tiny (Jy~b$^{-1}$)} & {\tiny (mas)} & 
{\tiny (mas)}  & \\
 \hline
6.035~GHz & RHC  & & & & \\
 \hline
$^*$1$_{RR}$  &  57.802 & 3.443 & -460.2 & -529.5 & z$_1$ \\
  2$_{RR}$  &  56.833 & 0.296 & -471.9 & -530.3 & z$_2$ \\
  3$_{RR}$  &  55.895 & 0.476 & -334.3 & -508.9 & z$_3$ \\
  4$_{RR}$  &  55.747 & 1.473 & -491.8 & -542.5 & z$_4$ \\
  5$_{RR}$  &  55.383 & 1.455 & -489.2 & -536.8 & z$_5$ \\
  6$_{RR}$  &  54.954 & 0.581 & -583.9 & -962.2 & z$_6$ \\
  7$_{RR}$  &  52.915 & 0.198 & -585.2 & -440.5 &{\em z$_7$$^{(\diamond)}$} \\
 \hline
 6.035~GHz & LHC  & & & & \\
 \hline
  1$_{LL}$  &  57.488 & 2.899 &  -460.2 & -529.5 & z$_1$ \\
  2$_{LL}$  &  56.451 & 0.255 &  -472.7 & -532.6 & z$_2$ \\
  3$_{LL}$  &  55.722 & 0.417 &  -332.6 & -509.6 & z$_3$ \\
  4$_{LL}$  &  55.656 & 0.895 &  -494.9 & -546.1 & z$_4$ \\
  5$_{LL}$  &  55.209 & 1.469 &  -490.9 & -539.5 & z$_5$ \\
  6$_{LL}$  &  54.725 & 0.463 &  -582.2 & -962.9 & z$_6$ \\
 {\em 7$_{LL}$} & { \em 52.590} & {\em 0.165} &{\em -575.1} & 
                                {\em -421.7} &{\em z$_7$$^{(\diamond)}$} \\
 \hline
 \end{tabular} \\
}
{\flushleft
\noindent $^*$: reference component

\noindent {\bf a:} the $\Delta \alpha$ and $\Delta \delta$ are relative to  
          the methanol reference maser spot position 
          RA$_{\rm J2000}$=$19^{\rm h}23^{\rm m}43\fs948$ 
          Dec$_{\rm J2000}$=14\degr30\arcmin34\farcs38

\noindent {\bf b:} Zeeman pair labelling \\

\noindent $\diamond$: Comp~7$_{LL}$ is quite faint and failed to pass the 
          3 consecutive channel criterion {\em (it only passed the 2 
          consecutive channel criterion)} but the Zeeman pattern is clearly 
          visible in the spectra (Fig.~\ref{fig: spectra OH 6035 z7}).
}
\end{table}

Following the selection criteria given in
section~\ref{sec: observations and data reduction},
seven maser components were identified in the RHC polarisation and 
six in the LHC. All of the 6 LHC components detected are associated with a 
RHC component in six Zeeman pairs at 6.035~GHz. 
These are presented in Table~\ref{Table: 6.035 GHz RHC and LHC comp info}. \\

\begin{figure}
      \epsfig{file=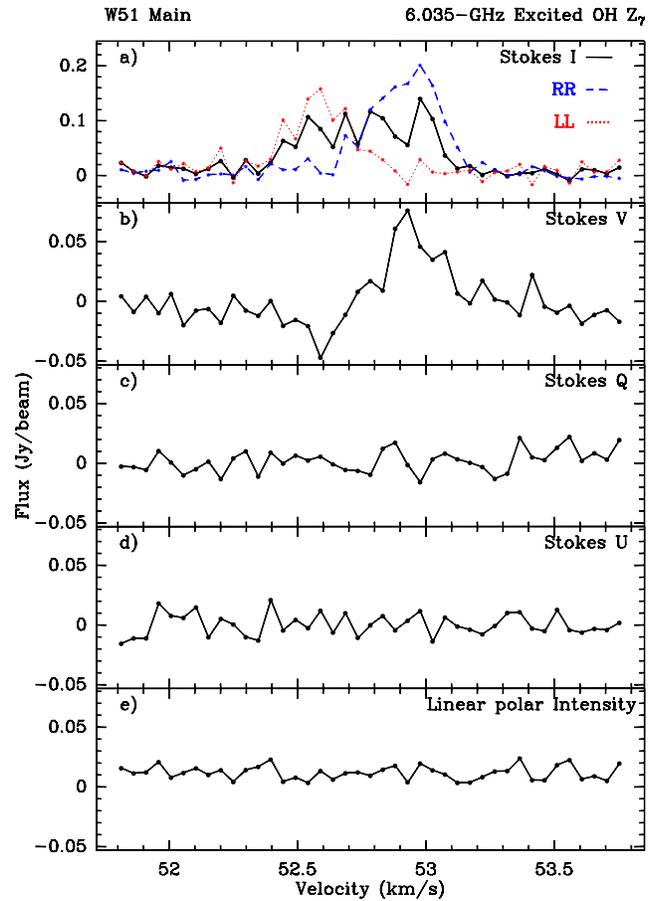,width=9.00cm}
\caption{Spectra of the Zeeman pair z$_7$ (6.035-GHz excited OH line). 
         {\bf a)} Stokes~$I$, RHC and LHC spectra. {\bf b)} Stokes~$V$. 
         {\bf c)} Stokes~$Q$. {\bf d)} Stokes~$U$. 
         {\bf e)} linear polarisation intensity $P$.}
\label{fig: spectra OH 6035 z7}
\end{figure}

The average Zeeman flux ratio Flux$_{\rm LL}$/Flux$_{\rm RR}$ is 0.83, ranging 
from 0.61 to 1.01. 
The seventh unpaired RHC component is the faintest component 
detected (less than 200~mJy). Taking into account the average flux ratio, the 
likelihood that this RHC component is also paired with a faint LHC 
counterpart that did not meet the 3 consecutive channel criterion was high.
And, indeed, the LHC component met the 3$\sigma$ threshold only over 2 
consecutive channels, but is clearly visible in the Stokes~$V$ spectrum with a 
peak intensity of 165~mJy (Fig~\ref{fig: spectra OH 6035 z7}). 
Consequently, 100\% of the emission detected in the excited OH line at 
6.035~GHz is part of a Zeeman pairing. \\

%%%%%%%%%%%%
%%%%%%%%%%%%
%%%%%%%%%%%%
%%%%%%%%%%%%

Table~\ref{table: 6.035 GHz demag info} presents, in decreasing velocity order, 
the characteristics of the Zeeman pairs identified, as follows: 
Column~1 gives the Zeeman pair label, Column~2 gives 
the demagnetized velocity (that is the best estimate of the actual velocity of 
the component), Columns~3 and 4 give the RA and Dec offset from the reference 
position used in the present work respectively, Columns~5 to 8 give the 
Stokes~$I$, $Q$, $U$ and $V$ intensity respectively, 
Column~9 gives the linear polarisation intensity $P$, 
Columns~10 to 12 give the circular, linear and total percentage of polarisation
($m_{\rm c}$, $m_{\rm l}$ and $m_{\rm t}$) respectively. In case of an 
elliptically polarised component (i.e., $m_l \ge 10$\%) Column~13 gives 
its associated polarisation angle $\chi$, Column~14 gives the velocity split 
($v_{\rm split}$=$v_{\rm (RHC)}$ - $v_{\rm (LHC)}$)
and finally the corresponding magnetic 
field strength B is given in Column~15. \\

%- - - - - - - - - - - - - - - - - - - - - - - - - - - - - - - - - - -
\begin{table*}
 \caption {\small 6.035-GHz Zeeman pair characteristics}
 \label{table: 6.035 GHz demag info}  
{\footnotesize
 \begin{tabular}{lrrrrrrrrrrrrll}
 \hline
  Z$^a$          & Vel$^b$       & $\Delta \alpha$$^c$ & $\Delta \delta$$^c$ &  
   I             &  Q            & U               & V               & 
   P             &  $m_{\rm c}$   & $m_{\rm l}$     & $m_{\rm t}$      & 
$\chi$           & $v_{\rm split}$& B   \\
                 & {\tiny (km~s$^{-1}$)}  & {\tiny (mas)} & {\tiny (mas)}  & 
{\tiny (Jy~b$^{-1}$)} & {\tiny (Jy~b$^{-1}$)} & {\tiny (Jy~b$^{-1}$)} & 
{\tiny (Jy~b$^{-1}$)} & 
{\tiny (Jy~b$^{-1}$)} & {\tiny (\%)} & {\tiny (\%)} & {\tiny (\%)} & 
{\tiny (\degr)}      & {\tiny (km~s$^{-1}$)} & {\tiny (mG)} \\

 \hline   
  z$_1$  &   57.645  & -460.2 & -529.5 & 1.8820 & -0.0040 & -0.0055 & 
 1.6450 & 0.0068 & 87.4 &  0.4  & 87.4 & ...  &  0.314  &   +5.57 \\
  z$_2$  &   56.642  & -472.3 & -531.4 & 0.1630 & -0.0124 &  0.0066 &  
 0.0960 & 0.0141 & 58.9 &  8.7  & 59.3 & ...  &  0.382  &   +6.77 \\
  z$_3$  &   55.808  & -333.5 & -509.3 & 0.3429 &  0.1006 &  0.0253 & 
 0.1740 & 0.1037 & 50.7 & 30.2  & 59.0 & +7.1 &  0.173  &   +3.07 \\
  z$_4$$^{(*)}$&   55.701  & -492.9 & -543.7  & 1.2290 & -0.0055 & -0.0029 & 
 0.2720 & 0.0062 & 22.1 &  0.5  & 22.1 & ...  &  0.091$^{(*)}$ & +1.61$^{(*)}$ \\
  z$_5$  &   55.296  & -490.1 & -538.0 & 1.2170 & -0.0071 & -0.0049 & 
-0.5690 & 0.0086 & 46.8 &  0.7  & 46.8 & ...   & 0.174  &   +3.08 \\
  z$_6$  &   54.839  & -583.1 & -962.5 & 0.4042 & -0.0068 & -0.0040 & 
-0.1990 & 0.0079 & 49.2 &  2.0  & 49.3 & ...   & 0.229  &   +4.06 \\
{\em z$_7$$^{(\diamond)}$} & {\em   52.759}  & {\em -580.3} & {\em -440.3} & 
    {\em 0.1400} & {\em -0.0094} & {\em -0.0300} & 
{\em 0.0762} & {\em 0.0095} & {\em 54.4}  & {\em 6.8} & {\em 54.8}  & ...  & 
{\em 0.340} &   {\em +6.02} \\  
\hline
 \end{tabular} \\

{\flushleft
\noindent {\bf a:} Zeeman pair labelling \\
\noindent {\bf b:} demagnetised velocity \\
\noindent {\bf c:} the $\Delta \alpha$ and $\Delta \delta$ 
          (Flux$_{RR}$*$\Delta_{{RR}}$+Flux$_{LL}$*$\Delta_{{LL}}$/(Flux$_{RR}$+Flux$_{LL}$))
          are with respect to  
          the methanol reference maser spot position 

          RA$_{\rm J2000}$=$19^{\rm h}23^{\rm m}43\fs948$ 
          Dec$_{\rm J2000}$=14\degr30\arcmin34\farcs38 \\
\noindent  {\bf d:} 
$v_{\rm split}$=$v_{\rm (RHC)}-v_{\rm (LHC)}$ \\

\noindent  {\bf *:} blending of many components on the line of sight 
           (cf. Fig.~\ref{fig: spectra OH 6035 z4 and z5}) \\
\noindent  {\bf $\diamond:$} the LL counterpart of this component is quite faint
and failed to pass the 3 consecutive channel criterion but the Zeeman pattern 
is clearly visible in the spectrum (Fig.~\ref{fig: spectra OH 6035 z7}). 
The Zeeman splitting has been consequently inferred from the Stokes~$V$ 
spectrum.  \\}
}
\end{table*}
%- - - - - - - - - - - - - - - - - - - - - - - - - - - - - - - - - - -

The seven Zeeman pairs identified in this transition were used to derive a 
magnetic field strength ranging from $+1.6$ to $+6.8$~mG, consistent with the 
previously published magnetic field strengths inferred from the OH 
ground-state lines in the region. 
Note that the weak magnetic field inferred from component z$_4$ ($+1.6$~mG) is 
due to blending of components in the line of sight making this measurement less 
reliable.

\begin{figure}
      \epsfig{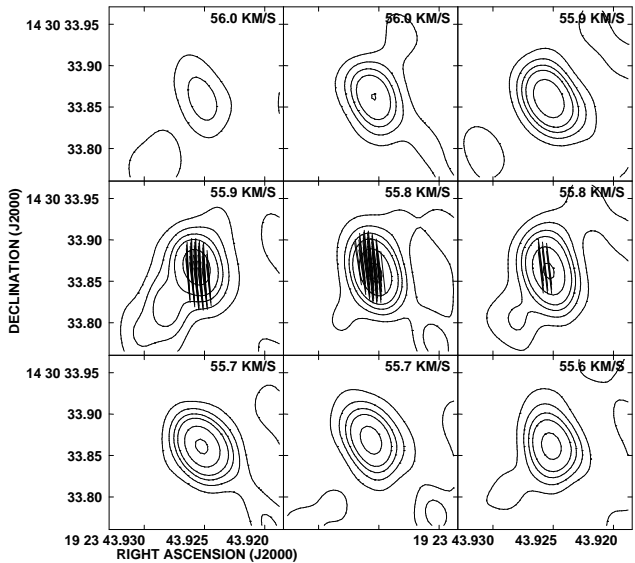}
\caption{Polarisation vector associated with the elliptical Zeeman component 
z$_3$.}
\label{fig: Linear polar info z3}
\end{figure}

The excited OH maser emission at 6.035~GHz shows a high degree of circular 
polarisation (typically $\ge 45$~\% and up to 87\%).
Note that only one Zeeman component (z$_3$) shows substantial linear 
polarisation (30\%) and has a polarisation angle  
$\chi=+7.1$\degr 
(Fig~\ref{fig: Linear polar info z3}). \\

%%%%%%%%%%%%
%%%%%%%%%%%%
%%%%%%%%%%%%
%%%%%%%%%%%%

%-  -  -  -  -  -  -  -  -  -  -  -  -  -  -  -  -  -  -  -  -  -  -  -
\subsection{Comparison with ground-state OH maser emission} 
\label{subsec: comparison with 18 cm OH}

Fish \& Reid (2007) present the distributions of the ground-state OH masers 
in the 1.665, 1.667 and 1.720~GHz transitions towards W51~Main/South 
obtained with the VLBA with a similar astrometric accuracy ($\sim$10~mas) as 
ours.

\begin{figure*}
\hspace*{-1cm} \epsfig{file=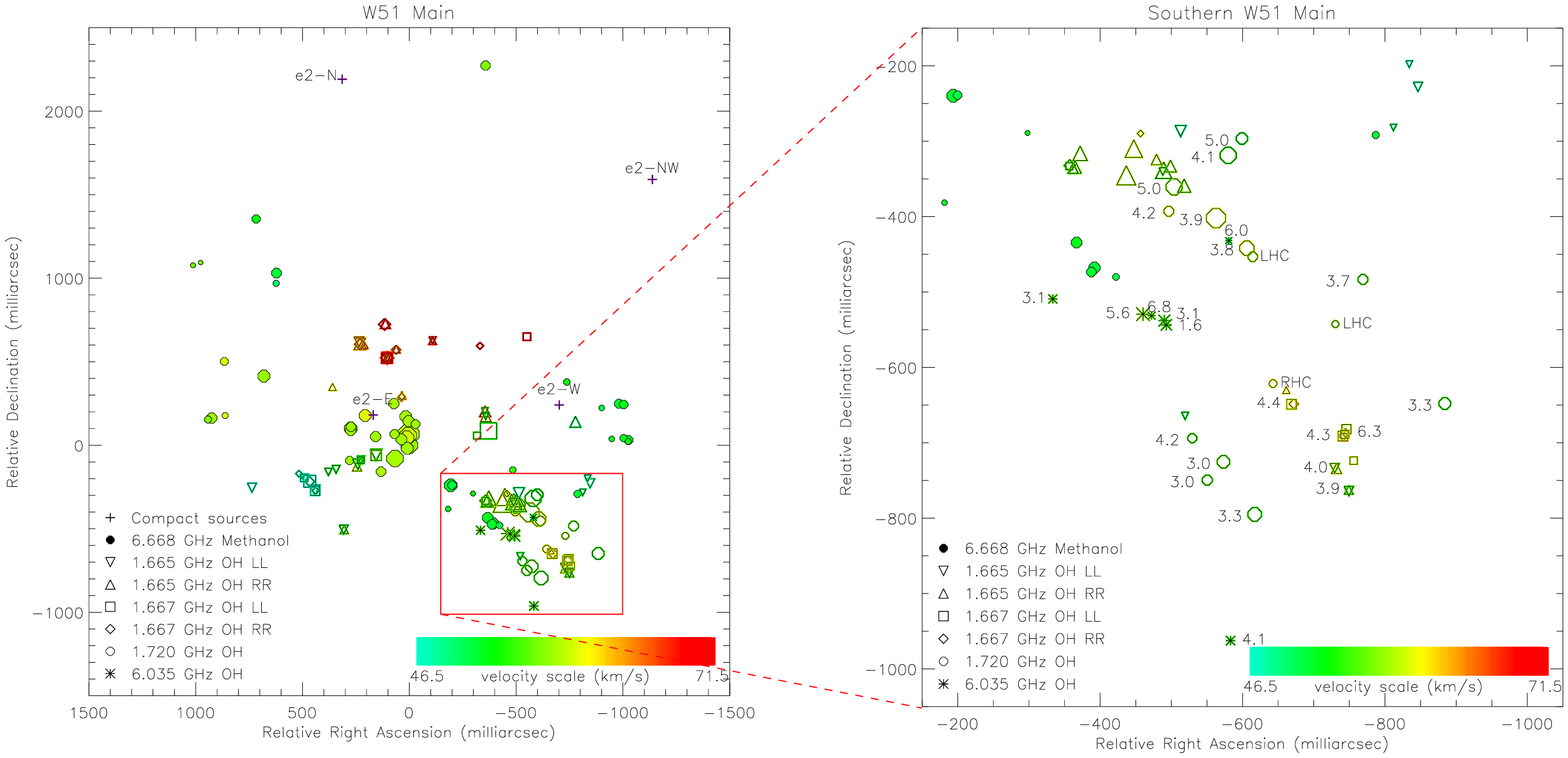,angle=-90,width=19.5cm}
\vspace{-2.5cm}
\caption{6.668-GHz methanol Stokes~$I$ components and 6.035~GHz 
         Zeeman components from this work with all the ground-state 
         OH maser components at 1.665 and 1.667~GHz in the RHC and LHC 
         polarisations and the 1.720~GHz Zeeman pairs and isolated RHC and 
         LHC components detected by Fish \& Reid (2007) in W51~Main.
         The size of the symbols is proportional to the log of the intensity.
         The crosses indicate the positions of the compact continuum sources 
         in the region
         (Shi et al. 2010a).
         Left panel: for the entire W51~Main region. Right panel: Magnification 
         of the region where 6.035-GHz excited OH maser emission is found.
         The magnetic field associated with all Zeeman pairs in all 
         the transitions is given by the side of each relevant component.}
\label{fig: OH ground state and excited W51 Main and Zoom}
\end{figure*}

Figure~\ref{fig: OH ground state and excited W51 Main and Zoom} presents the 
6.668-GHz methanol components and 6.035~GHz components we detected along with  
all the ground-state OH maser components at 1.665, 1.667 and 1.720~GHz  
detected by Fish \& Reid (2007) in W51~Main.
The ground-state OH maser emission does not alter the overall orientation 
along a P.A. $\simeq 70\degr$ and confirms the lanes/gaps devoid of maser 
emission clearly observed in the methanol maser distribution. 
Note that the magnetic field strengths given in 
Figure~\ref{fig: OH ground state and excited W51 Main and Zoom} for 1.720~GHz 
differ by a factor of 2 to those inferred by Fish \& Reid (2007) as calculated 
from the Zeeman splitting coefficient given by Davies (1974) and implying 
potential blending of $\sigma$ components. Fish \& Reid (2007) assumed that all 
detected 1720-MHz components were of $\sigma^{\pm 1}$ type, for which the 
Zeeman splitting coefficient is half that of Davies (1974).
The magnetic field strengths inferred from the ground-state OH lines by 
Fish \& Reid (2007) and from the excited OH line at 6.035~GHz (this work) are 
in general agreement. \\

%%%%%%%%%%%%
%%%%%%%%%%%%
%%%%%%%%%%%%
%%%%%%%%%%%%

%%%%%%%%%%%%
%%%%%%%%%%%%
\begin{figure}
      \epsfig{file=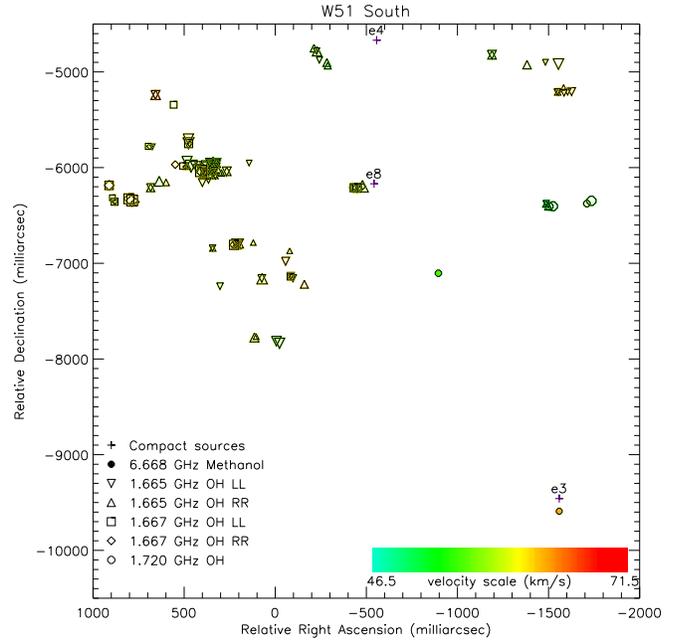,width=8.65cm}
% \vspace{-2.5cm}
\caption{6.668-GHz methanol Stokes~$I$ components from this work with all the 
         ground-state OH maser components at 1.665 and 1.667~GHz in the RHC 
         and LHC polarisations and the 1.720~GHz Zeeman pairs and isolated RHC 
         and LHC components detected by Fish \& Reid (2007) in W51~South.
         The size of the symbols is proportional to the log of the intensity.
         The crosses indicate the positions of the compact continuum sources 
         in the region
         (Gaume et al. 1993, Zhang \& Ho 1997).
         }
\label{fig: OH ground state and excited W51 South}
\end{figure}
%%%%%%%%%%%%
%%%%%%%%%%%%

Considering both the methanol and OH in W51~Main, the overall maser emission 
seems to be centered on W51e2-E. In the northern region, covering the area
$\Delta \alpha$=[+1000,+500]~mas $\Delta \delta$=[+100,+1400]~mas only methanol 
maser emission is found. In the regions centered on W51e-E and south-west of it,
maser emission of both species is found, but a close look shows that methanol 
and OH maser components are not overlaping. \\

Figure~\ref{fig: OH ground state and excited W51 South} presents the 
6.668-GHz methanol components we detected together with all the ground-state 
OH maser components detected by Fish \& Reid (2007) in W51~South. 
In this region, the dichotomy between the OH and methanol maser distributions 
is even more pronounced. Methanol maser emission is very scarce and found in 
very clear distinct regions. We also note the total absence of 6.035~GHz 
emission and the scarcity of 1.720~GHz emission.

%----------------------------------------------------------------------
\section{Modelling procedure}
\label{sec: modelling}

We have carried out a new parameter-space search for inversion in OH lines, 
up to, and including the 13-GHz transition in the $^2\Pi_{3/2}, J=7/2$ 
rotation state. The models used the Accelerated Lambda Iteration (ALI) method 
to solve the coupled radiation transfer and statistical steady-state problems.
The radiative transfer included only the far-infrared (FIR) transitions of OH, 
so it predicts only maser optical depths, without saturation or competitive 
propagation effects.

The model included the lowest 36 hyperfine levels of OH, coupled by 137 
radiatively-allowed transitions (including the potentially maser-active 
hyperfine lines). Collisions of OH with both ortho- and para-H$_2$ were 
included, using the rate-coefficient from 
Offer \& van~Dishoeck (1992). The ortho- and para-H$_2$ 
ratio was 3 in the low-temperature limit, and at all model kinetic 
temperatures controlled by the Boltzmann distribution amongst the rotational
levels of H$_2$. A semi-infinite slab geometry was used with 85 
logarithmically-spaced slabs, covering a total depth of $3 \times 10^{15}$~cm.
The outer, (nearer to observer) $2 \times 10^{15}$~cm, covering 80 slabs, 
contained OH, and had a fixed set of physical conditions 
(except the bulk velocity) for each model. The 5 slabs that are most remote 
from the observer behave as an 
optically thick boundary, with an exponentially increasing dust abundance, 
and an exponentially decreasing OH abundance. The parameter space ranges 
covered by the model are listed in Table~\ref{table: OH model parameter space}.
All models used an OH abundance of $2 \times 10^{-7}$ w.r.t. H$_2$.

%- - - - - - - - - - - - - - - - - - - - - - - - - - - - - - - - - - -
\begin{table}
 \caption {\small Model parameter ranges}
 \label{table: OH model parameter space}  
{\footnotesize
 \begin{tabular}{lll}
 \hline
 Parameter       & Minimum value   & Maximum value   \\ 
 \hline
 Kinetic Temp. & 30K             & 250K            \\
 Dust Temp.      & 10K             & 300K            \\
 n(H$_2$)        & 10$^6$cm$^{-3}$  & 10$^9$cm$^{-3}$ \\   
 $v_{microturb.}$ & 0~km~s$^{-1}$ & 0~km~s$^{-1}$ \\
 $\Delta$v       & $-10$~km~s$^{-1}$ & $+10$~km~s$^{-1}$ \\
 \hline
 \end{tabular} \\
}
{\flushleft
\noindent {Note: All the above parameters were considered to be independent}.
}
\end{table}
%- - - - - - - - - - - - - - - - - - - - - - - - - - - - - - - - - - -

All the model parameters were considered to be independent.
The dust parameters, for a mixture of carbonaceous and silicate dust, were 
taken from Draine \& Lee (1984), and this was admixed with the 
OH-bearing slabs at an abundance of 1~per cent by mass. \\

%%%%%%%%%%%%
%%%%%%%%%%%%
\begin{figure*}
\epsfig{file=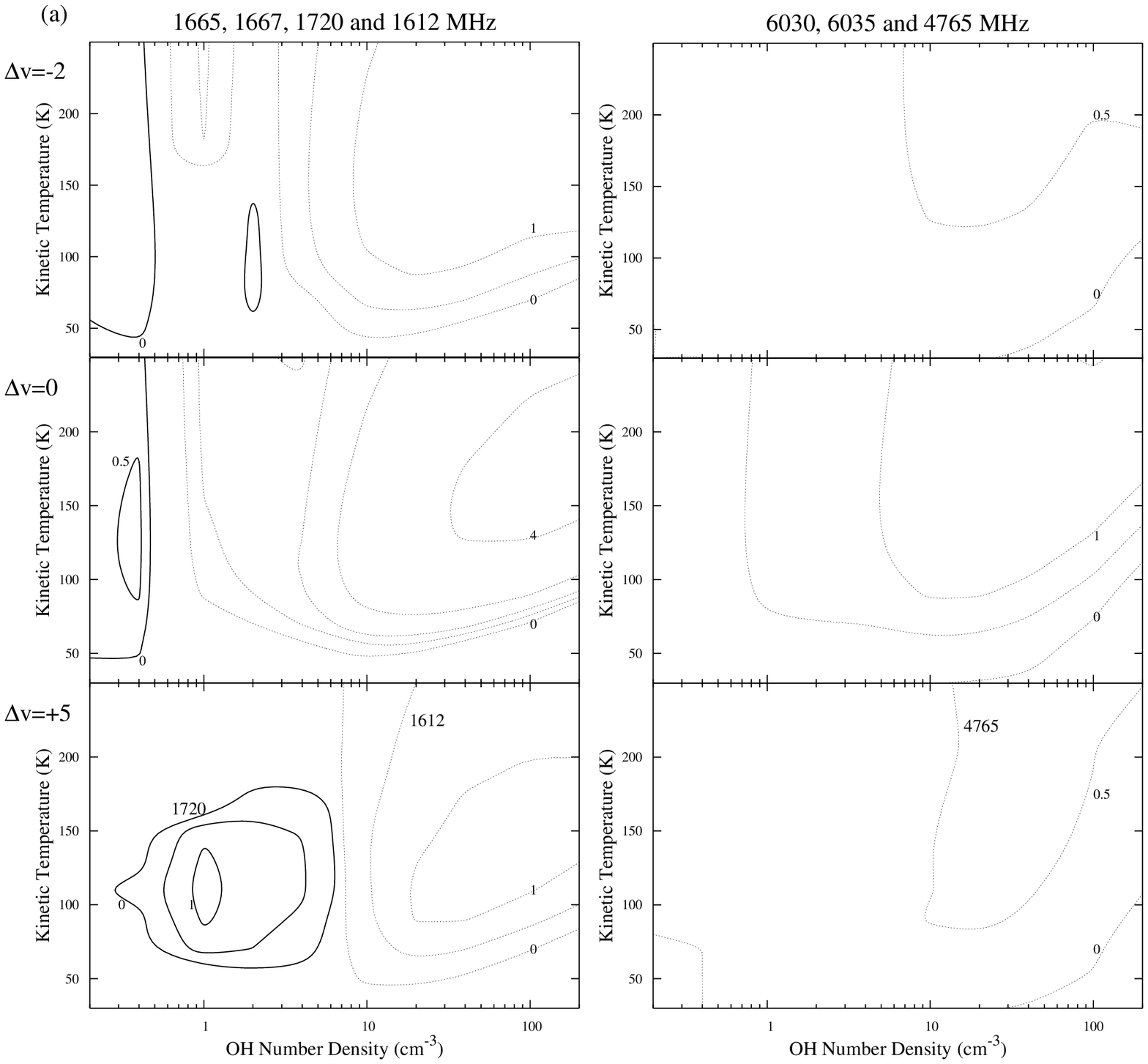,angle=0,width=12.25cm}
\epsfig{file=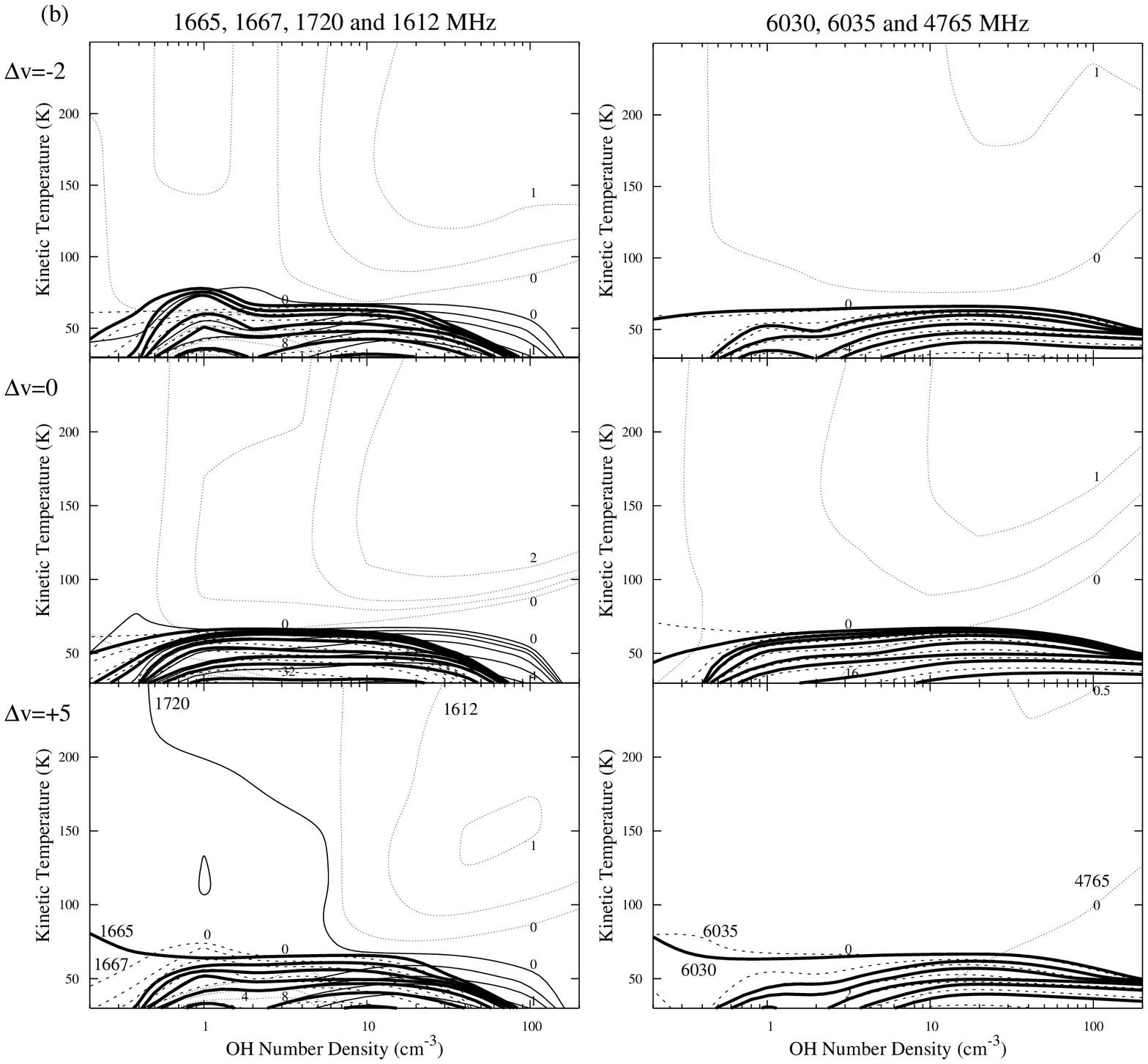,angle=0,width=12.25cm}
\vspace*{-0.3cm}
\caption{Contour plots of maser optical depths for selected OH transitions. 
{\bf a},~{\bf b},~{\bf c} show the inversion present at dust temperature 
T$_{\rm d}$ of 10, 72 and 134~K, respectively. 
The line styles for the left-hand side plots are heavy solid (1.665~GHz), 
dashed (1.667~GHz), light solid (1.720~GHz) and complex (1.612~GHz). 
The line styles for the right-hand side plots are heavy solid (6.030~GHz), 
dashed (6.035~GHz) and complex (4.765~GHz). Contour levels are at 
$\tau$=0, 0.5, 1, 2 and then doubling up to $\tau$=64.
All models used an OH abundance of $2 \times 10^{-7}$ w.r.t. H$_2$.
}
\label{fig: OH Modelling outcome}
\end{figure*}
%%%%%%%%%%%%
%%%%%%%%%%%%

% % % % % % % % % % % % 
        \addtocounter{figure}{-1}%
% % % % % % % % % % % %
%%%%%%%%%%%%
%%%%%%%%%%%%
\begin{figure*}
\epsfig{file=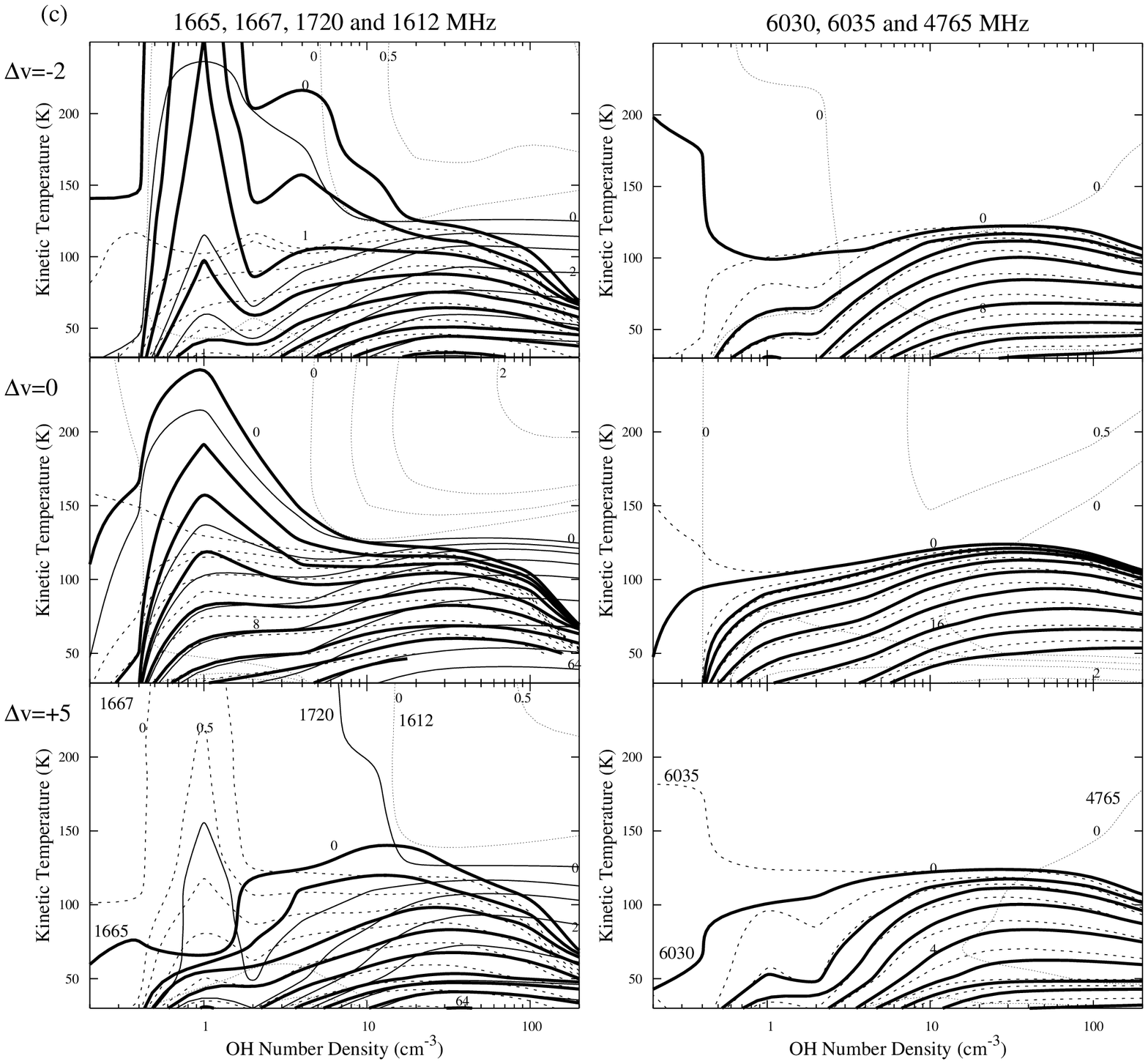,angle=0,width=12.25cm}
%% \vspace{-2.5cm}
\caption{\it continued} 
\end{figure*}
%%%%%%%%%%%%
%%%%%%%%%%%%

In Fig.~\ref{fig: OH Modelling outcome}, we plot contours of the maser 
optical depth $\tau$, for notional propagation perpendicular to the slabs, and
$\tau= {\sum_{i}} \gamma_{i} \delta Z_{i}$, 
over the 80 non-boundary slabs (i=1-80), 
where $\gamma_{i}$ is the maser gain coefficient for the slab~i. Contour levels
are at $\tau$=0, 0.5, 1, 2 and then doubling up to $\tau$=64.
Fig.~\ref{fig: OH Modelling outcome}a,~b,~c show the inversion present at dust 
temperature T$_{\rm d}$ of 10, 72 and 134~K, respectively. 
At T$_{\rm d}$=10~K (Fig.~\ref{fig: OH Modelling outcome}a) we find that the 
inversions are present only for the satellite lines, and that they are weak. 
For the ground-state lines, the inversion zones in the OH number density, 
kinetic temperature (n(OH), T$_{K}$) plane for 1.612 and 1.720~GHz are 
distinct and significantly modified by the velocity shift, via the effects of 
FIR line overlaps. At T$_{\rm d}$=72~K (Fig.~\ref{fig: OH Modelling outcome}b), 
the main lines appear in both the $^2 \Pi_{3/2}$ J=3/2 
(1.665 and 1.667~GHz)
and $^2 \Pi_{3/2}$ J=5/2 
(6.031 and 6.035~GHz)
$\Lambda$-doublets at low T$_{K}$. 
In general, there is little or no main-line inversion when 
T$_{K} >$ T$_{d}$, which reflects the radiation-dominated pumping of these 
lines (Gray 2007). Peak inversions for the 6-GHz maser lines lie at larger 
values of n(OH) than for the 1.720-GHz line. At T$_{\rm d}$=134~K 
(Fig.~\ref{fig: OH Modelling outcome}c), we can see a low density inversion 
region for the 1.665 and 1.667~GHz lines that is line-overlap driven. This 
region is absent for the 6~GHz transitions. These conclusions are in broad 
agreement with earlier work (Gray, Doel \& Field 1991; Gray et al. 1992), but 
we note that these earlier parameter-space searches used the LVG 
approximation, and could therefore not study the important static 
($\Delta$v=0)
models shown in the middle row in 
Fig.~\ref{fig: OH Modelling outcome}a,b,c.

%----------------------------------------------------------------------
\section{Discussion}
\label{sec: discussion}

\subsection{OH and CH$_3$OH masers}
\label{Discussion:Oh and methanol masers}

Although not detected in these observations, maser emission in the 6.031-GHz
excited transition of OH was reported by Rickard et al. (1975). The velocity
range of their sole 6.031-GHz component coincides with
that of the faint 6.035-GHz Zeeman pair $z_7$ here, which was 
also the strongest component detected by Rickard et al. (1975). 
On the other hand, the strongest 6.035-GHz component detected here, $z_1$, 
was not detected by Rickard et al. in the 1970's. 
Such variability in the intensity of the maser components themselves, but also 
in their intensity ratios, is likely to account for the non-detection of 
6.031~GHz emission here. \\

Overall the bulk of the methanol maser emission, both in terms of flux
and number of components, is associated with W51 Main. Only two weak,
isolated components are seen in W51~South, most likely associated with
the e3 and e8 sources. Although the work of Fish \& Reid (2007) shows
that there is significant ground state OH emission from both Main and
South, with indeed the brightest OH component being in South, excited
OH emission is only detected towards Main. 

As noted by Etoka et al. (2005) in the case of the SFR 
W3(OH), the present astrometric study towards W51~Main/South confirms that 
associations of individual OH and methanol maser components are rare. 
Despite all the components identified in the rich maser environment of  
W51~Main, no methanol and OH maser component overlap is
found. 
The minimal separation between a ground-state OH and a 
methanol component is 27~mas ($\sim150$~AU), rising to 60~mas ($\sim300$~AU) 
between an excited OH and a methanol component. This is suggestive of
local variations in the abundance of the species and that
both species are found in closely associated, but distinct, pockets.
Similarly, even though 6.035-GHz excited-OH and ground-state OH maser
components are found in similar areas in W51~Main with similar magnetic field
strength, they too do not show any overlap when observed at high spatial
resolution.

Based on the present OH modelling results, and those of Gray et al. (1992) and 
Cragg et al. (2002), the total absence of 6.035-GHz emission and the scarcity 
of 1.720~GHz emission in W51~South, is suggestive of a lower density in 
W51~South than in W51~Main. \\

The physical conditions under which 
6.668-GHz 
methanol masers
(and 12.2-GHz methanol masers) form appears to be very similar to those that 
support the OH masers in almost all respects (Cragg, Sobolev \& Godfrey 2005). 
In particular, maser optical depths are rather insensitive to the dust 
temperature provided the criterion $T_d > T_k$ is met, and they decay with 
increasing $T_k$ for a given dust temperature. Overall gas density also 
appears unable to differentiate between the OH and methanol pumping regions.

The parameter that does appear to distinguish the methanol inversion region 
in parameter space from that of OH is the specific column density 
(Cragg et al. 2005), that is the number density of the active molecule 
divided by the velocity gradient in the medium. Although this model cannot 
be applied to the middle row of Figure~\ref{fig: OH Modelling outcome}, 
 it can be computed for the other two rows, where the velocity 
gradient is $5\times 10^{-11} \Delta v$\,s$^{-1}$. For conditions that
strongly favour 1.7-GHz to 6.0-GHz OH masers, we are restricted to number 
densities of OH below $\sim$10\,cm$^{-3}$, or to specific column densities of 
less than $2\times 10^{11}/\Delta v$\,s\,cm$^{-3}$. The models in Cragg et al. 
require specific optical depths of methanol that are typically a factor of 
ten larger. The controlling factor in which species of maser appears from a 
location at VLBI resolution therefore appears to be either abundance of the
active molecule, or the local velocity gradient, with more quiescent gas 
favouring methanol over OH masers.

%- - - - - - - - - - - - - - - - - - - - - - - - - - - - - - - - - - - - - -
\subsection{Distribution of masers around W51e2}
\label{Discussion: W51e2}

%%%%%%%%%%%%
%%%%%%%%%%%%
\begin{figure*}
\hspace*{-1cm} \epsfig{file=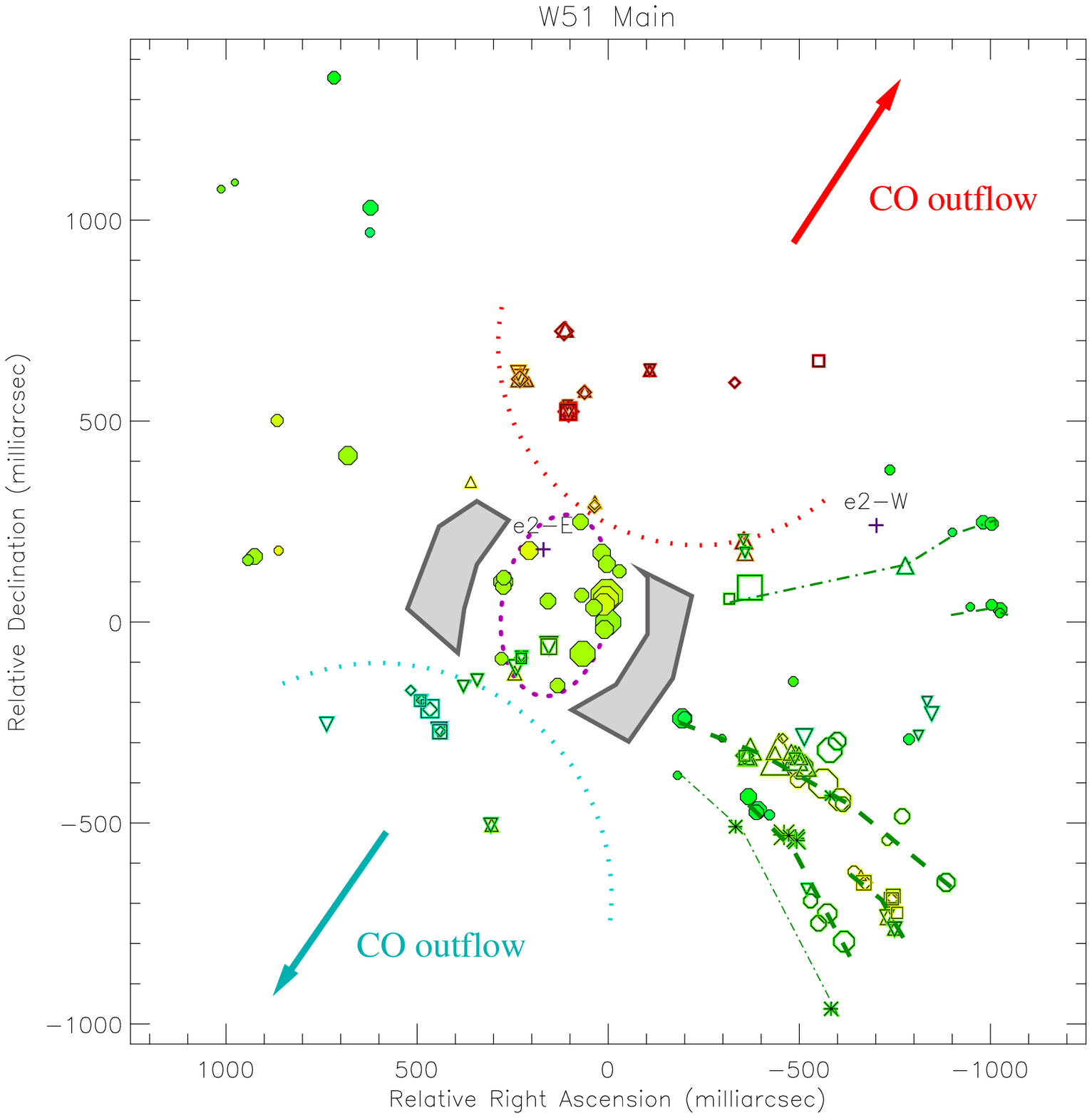,angle=-90,width=19.5cm}
\vspace{-1.2cm}
\caption{Magnification on the masers around e2-E and e2-W presenting the 
        various spatial-kinematic components identified. 
        The orientation of the CO outflow observed by Shi 
        et al. (2010b) is indicated along with the 2 lanes/gaps 
        (grey shaded polygons) devoid of methanol and OH maser emission.
      We refer to Fig.~\ref{fig: OH ground state and excited W51 Main and Zoom} 
        for the legend.}
\label{fig: OH ground state and excited W51 Main entire structure Zoom}
\end{figure*}
%%%%%%%%%%%%
%%%%%%%%%%%%

Fig.~\ref{fig: OH ground state and excited W51 Main and Zoom} shows that there 
are several distinct spatial-kinematic components in the region that are 
identified in 
Fig.~\ref{fig: OH ground state and excited W51 Main entire structure Zoom} and 
discussed here. 
The most extreme velocity maser components
(within the velocity range $\sim$[+46.5,+50]~km~s$^{-1}$ 
and $\sim$[+66,+71.5]~km~s$^{-1}$ for the bluest and reddest components 
respectively),
which are all ground state OH
masers, lie close to a line through W51e2-E at a P.A.~$\sim150^\circ$ 
with the redshifted components north of the source and blueshifted
components offset by a similar angular distance ($\sim400$~mas) south of the
source. The location of these components as well as their velocities indicate 
that they are associated with the outflow from the source which has been 
imaged by Shi et al. (2010b). 
Compared with these extreme velocity OH maser components, the methanol maser
components, all the ground-state OH maser components at 1.720~GHz and the 
excited OH maser components at 6.035~GHz along with the ground-state main-line
components found in the neighbouring areas (typically $\Delta \alpha \le
0$~mas and $\Delta \delta \le 200$~mas) have intermediate velocities. \\

The methanol masers are predominantly clustered within about {1\arcsec} of
W51e2-E. Three major groupings separated by regions without any
methanol components can be identified 
(Fig.~\ref{fig: OH ground state and excited W51 Main entire structure Zoom}): 
a number of spots east and north east of W51e2-E, a ring- or arc-like 
structure the centre of which is within about 200~mas of e2-E, and the 
remaining components which extend further to the west and southwest with a 
suggestion of falling into elongated structures (filaments) extending to the 
west or southwest and pointing back towards W51e2-E.

The majority of the excited OH masers are located at the south western edge 
of the methanol maser distribution, as a part of one of the most 
prominent filaments, extended by ground state OH masers further to the south 
west. The majority of the ground state OH masers trace a second, approximately 
parallel, filament about 100~mas north of the methanol-excited OH filament. 
This second prominent filament also has two methanol maser components at the
eastern end, the end closest to W51e2-E.

The spatial segregation of the maser species around W51e2-E as well as the
lack of overlap of components from different species, plus the extreme
difference in the richness of the methanol maser emission towards W51 Main and
South, both point to significant variations in the physical and chemical
conditions in the region as has been noted previously by Fish \& Reid (2007).\\

%- - - - - - - - - - - - - - - - - - - - - - - - - - - - - - - - - - - - - -
\subsection{Polarisation and magnetic field}
\label{Discussion: B field structure}

The magnetic field strengths inferred here from the 6.035-GHz excited OH
Zeeman components range from $\sim+2$~mG to $\sim+7$~mG 
(Table~\ref{table: 6.035 GHz demag info}), consistent with the 4.1~mG 
reported by Rickard et al. (1975) from the 6.031-GHz excited OH transition 
and inferred from the ground state OH transitions as discussed in 
section~\ref{subsec: comparison with 18 cm OH}. \\

Polarisation angle measurements have been obtained towards W51e2 by 
Lai et al. (2001) from BIMA 1.3-mm dust continuum observations, further 
analysed by Tang et al. (2009). The polarisation angle they measure in the 
region where the elliptical OH Zeeman component z$_3$ emanates from, is in 
agreement with the polarisation angle we measure. 
Tang et al. (2009) conclude from the analysis of the overall polarisation 
angle distribution for the BIMA observations at 1.3-mm that the magnetic field 
has an hourglass morphology pinched along a plane consistent with the
H53$\alpha$ accretion flow proposed by Keto \& Klaassen (2008).
The orientation of the proposed flow (P.A. $\simeq 60\degr$) is consistent with
the distribution of the intermediate velocity maser population 
(P.A. $\simeq 70\degr$). We also note that the orientation of the 
filamentary structures observed in the southwestern regions 
(cf. section~\ref{Discussion: W51e2}) seems also in agreement with the 
magnetic field line directions inferred by Tang et al. (2008) for this region 
and given in their $B$-field maps (their Fig.~5). 
This seems to indicate that the maser features are 
likely to be part of a flow in which the dynamics are magnetically dominated, 
so that motion along the field lines is significantly easier than accross them.
The filamentary structure orientation can be explained if the maser spots trace 
the magnetic field lines.
\\

%- - - - - - - - - - - - - - - - - - - - - - - - - - - - - - - - - - - - - - 
\subsection{Interpretation and scenarios}
\label{Discussion: Scenarios}

Fig.~\ref{fig: Methanol components W51Main}b presents the velocity 
distribution of the methanol maser spots versus radial distance. The figure 
shows that the red- and blue-shifted maser components are diverging from a 
common point with an expansion velocity of $\simeq$5~km~s$^{-1}$. 
This, added to the general distribution of the methanol maser components in the 
plane of the sky (Fig.~\ref{fig: Methanol components W51Main}a \& 
Fig.~\ref{fig: OH ground state and excited W51 Main entire structure Zoom}), 
indicates the presence of a structure with a wide opening angle centred on 
W51e2. \\

We can estimate roughly the total mass traced by the structure from the 
material that would be contained by the enclosing sphere.
With a total extent of $\sim$1.5{\arcsec} and assuming 
n(H$_2$)=$5 \times 10^7$cm$^{-3}$ 
from the OH modelling of the region, the upper limit for the total mass is 
90~M$_\odot$. An estimate of the lower limit can be obtained by considering 
only the volume strictly traced by the methanol masers. Considering the 
$\sim45 \degr$ opening angle of the structure, translating into a filling 
factor of 0.25, gives us 22.5~M$_\odot$. \\

\noindent
{\bf Scenario 1: Outflow} \\
\indent
One possible interpretation for the wide-opening angle structure observed is 
the presence of an outflow.

According to Churchwell (2002), typically an outflow from a high-mass young 
stellar object (HMYSO; 
i.e., L$_{bol}> 10^3$ L$_\odot$) lives $\sim 10^4$~yr with a mass 
outflow rate of $\sim 10^{-3}$~M$_\odot$~yr$^{-1}$. With an expansion 
velocity of 5~km~s$^{-1}$, only $\sim 5 \times 10^{3}$~year would be needed 
to reach {1\arcsec}, corresponding to the extent of the red- and blue shifted 
lobe observed in Fig.~\ref{fig: Methanol components W51Main}a.

Considering this time scale and the lower and higher total mass limits inferred
here, this would lead to an outflow rate of 
$\simeq 2\times 10^{-2} \rightarrow 4\times 10^{-3}$~M$_\odot$~yr$^{-1}$. This
is quite similar to the mass outflow rate inferred by Shi et al. (2010b) for 
the CO outflow observed perpendicular to the wide-angle structure traced by the 
masers.

In this scenario, the 2 gaps observed in the distribution of the masers could 
be explained by 2 episodic events. The ring-like structure closest to 
e2-E, well defined by an ellipse of long-axis $\sim 220$~mas, would trace an  
outflow event $\sim 1.1 \times 10^3$~yr old only, assuming no acceleration or 
deceleration. \\

\noindent
{\bf Scenario 2: Accreting flow} \\
\indent
The northeast-southwest distribution of the masers is close to
perpendicular to the axis of the well collimated outflow of material
imaged in CO by Shi et al. (2010b). This suggests that alternatively the 
masers could be tracing material which is part of the static or infalling 
envelope around the forming star.

The excitation models show that both OH and methanol maser emission
are suppressed by higher gas temperatures. This could provide an
explanation for the gaps which separate the two outer groups of
methanol masers from the central cluster. Closer to the central source
the gas is likely to be hotter and so maser emission is suppressed,
even if there is a high abundance of methanol and OH. Alternatively,
the gaps could reflect regions where the infall velocity increases,
destroying the velocity coherent path along the line of sight necessary
for maser amplification.

The methanol masers closest to e2-E, forming a compact ring-like structure, 
likely trace a distinct physical component. Similar structures are
seen by Bartkiewicz et al. (2009) towards some 29\% of the 6.7-GHz
methanol maser sources they studied. They interpreted these as a
result of the masers being associated with a dense disk or torus
around the central source, an interpretation which could also be
applied to the masers seen here. The fact that these masers exhibit  
similar red-shifted velocities as the masers located $\sim 1\arcsec$ 
north-east of e2-E could be due to radiative transfer effects in the 
infalling envelope.

%- - - - - - - - - - - - - - - - - - - - - - - - - - - - - - - - - - - - - -
\subsection{Stage of evolution and properties of the sources}
\label{Discussion: Stage of Evolution}

Masers have been proposed as tracers of the evolutionary stage of 
HMYSOs which excite them. For example Breen et al. (2010) propose a scheme 
in which sources evolve from exciting only
methanol masers to having both methanol and OH masers while their
UCHII region grows. Subsequently, the methanol maser emission ceases
leaving OH masers associated with the star and its UCHII region.
Within such a scheme these observations of W51~Main and South suggest
that the methanol-poor South region is more evolved than the much
more methanol-rich Main region. In South, the ground-state OH masers
appear primarily associated with UCHII region e8. Both the methanol
and OH masers in Main are associated with the presumably younger dust
continuum dominated source e2-E with little evidence that indicates any
of the masers are associated with the UCHII region e2-W. 

Shi et al (2010a) propose that there is either a single O~protostar or a 
cluster of B stars at the centre of W51e-E. 
The results of Keto \& Klaassen (2008) and Shi et al. (2010b) seems to indicate 
that there are 2 possible HMYSO in the W51e2 region: W51e2-E and W51e2-W.
Shi et al. (2010b) reach the conclusion that W51e2-W is somewhat older than 
W51e2-E and was formerly the active centre of the region but has now ceased to 
accrete and that it is now W51e2-E which dominates the gas accretion in the 
W51e2 molecular core. The presence of the methanol wide-opening angle 
structure with no UCHII region association points to the existence of a 
O-protostar in the early stage of evolution in the cluster. \\

Interferometric observations of methyl cyanide towards the W51 Main
and South region by Remijan et al. (2004) identified hot gas
associated with e2 and e1, although the observations had insuffient
resolution to separate the components of e2.  At the $2\arcsec-3\arcsec$
resolution of the observations the e2 and e1 regions were resolved
with the lines peaking at a velocity of 56~km~s$^{-1}$ towards e1 and 
53~km~s$^{-1}$ towards e2. Many of the lines are optically thick but 
correcting for this, the analysis found e1 and e2 to have similar 
temperatures ($\sim130$K), methyl cyanide column densities
(few$\times10^{16}$cm$^{-2}$) and hydrogen volume densities
($\sim5\times10^5$cm$^{-3}$).  However, contrary to what might be
expected on the basis of these methyl cyanide results, there are no
maser components associated with the e1 source. The masers in the
south are associated with e8, except for the lone methanol component
associated with e3. Whether this difference between the masers and the
hot methyl cyanide reflects a difference in the evolutionary status of
the e1 and e8 sources is unclear, but higher resolution observations
of both the thermal line emission and the dust continuum emission from
this region would provide important insights into this question.

%- - - - - - - - - - - - - - - - - - - - - - - - - - - - - - - - - - -
\subsection{Comparison with similar SFR complexes}
\label{Discussion: comparison with other SRFs}

Studies of the 6-GHz excited-state OH and 6.668-GHz methanol maser
emission with similar astrometric accuracy to those presented
here have also been performed towards the SFR complexes
W3(OH) (Etoka et al. 2005) and ON1 (Green et al. 2007). The
results in ON1 are remarkably similar to W51. In particular, in
ON1 there is no association between 6-GHz excited-state OH and
methanol masers with the closest components of each species being
separated by $\sim$23~mas. In addition, the 6-GHz excited-state OH
and methanol masers form a linear distribution. However this
structure is systematically offset (by $\sim$60-70~mas) from another
linear distribution traced by ground-state OH masers.  This
offset could result from the proper motions of the masers in the
9 years between the observations of the ground-state
OH (Nammahachak et al. 2006) and the methanol and 6-GHz excited-state OH.
Nonetheless, this filamentary distribution is strikingly
reminiscent of that observed here in the south-southwest part of
W51 Main. In addition, the magnetic field measured in ON1 is
similar in strength to that in W51. These similarities suggest
a similar mechanism/environment is observed in both SFR complexes. \\

In contrast, the maser distribution in W3(OH) is quite 
different from W51 or ON1. Even though long filamentary structures are
observed, the most striking one being the large-extended filament
traced both by 6.668-GHz methanol and 4.765-GHz excited-OH maser
emission (Harvey-Smith \& Cohen 2006), there is no clear apparent pattern 
nor segregation of the different species in the area probed by the maser 
components.

%----------------------------------------------------------------------
\section{Summary and conclusion}
\label{sec: conclusion}

We have presented MERLIN astrometric observations towards W51~Main
and South of the Class~II methanol maser emission at 6.668~GHz and the 
excited OH maser emission at 6.035~GHz. The 6-GHz maser distributions have been 
aligned with those of the ground-state OH maser transitions at 1.665, 1.667 and 
1.720~GHz from Fish \& Reid (2007). Although Main and South have similar 
number of OH ground-state maser components with the strongest component in 
South, the bulk of the methanol 6.668~GHz maser emission, and 
all of the excited OH 6.035~GHz maser emission are found to be 
associated with e2 in W51~Main. Only two faint methanol maser spots 
are found in W51~South, probably associated with e3 and e8.
Modelling implies n(H$_2$)$ \simeq 5 \times 10^7$cm$^{-3}$, a dust temperature 
30~K$\le $~T$_{d} \le$130~K and a kinetic temperature T$_K \le$T$_{d}$
as reasonable physical conditions for Main with a possibly lower density 
for South. 

At this high spatial resolution
(better than 15~mas), no overlapping of OH and methanol maser spots or 
6.035-GHz excited-OH and ground-state OH maser components are found even in the
most crowded areas. 

The alignment of the 18-cm ground-state masers and the 5-cm methanol and 
excited OH masers also reveal that the extreme velocity main-line masers 
found $\simeq$400~mas north and 
south of e2-E (either side of the methanol ring-like struture) are associated 
with the well-known outflow observed in CO by Shi et al. (2010b).
 
The masers in the south-southwest region of W51~Main clearly exhibit a 
filamentary structure. Those filaments extend west or southwest and point back 
towards e2-E. There are 2 prominent filaments, one of which contains the bulk 
of the 6.035-GHz maser spots, located at the southwest edge of the methanol 
maser spots and extended by 1.720-GHz maser spots. The magnetic 
strength and direction inferred from the excited OH maser emission at 
6.035~GHz is in agreement with former measurements in the region. Also, 
our measured polarisation angle is in agreement with BIMA 1.3-mm 
observations, interpreted by Tang et al. (2009) as the signature of an 
hourglass magnetic field structure suggesting that the maser spots trace the 
magnetic field lines.

A close inspection of the methanol masers revealed a wide-opening angle 
structure centred on e2-E, roughly aligned on a P.A.$\sim 150\degr$,
that is roughly perpendicular to the CO outflow,
and showing a clear velocity coherence. We estimated that the mass  
in this structure is between 22.5 and 90~M$_\odot$. The two possible 
interpretations of this structure are the signature of {\bf (1)} an outflow 
showing episodic events of $\sim5 \times 10^3$~yr for the older event 
and $\sim1 \times 10^3$~yr for the younger one, assuming an outflow velocity 
of $\sim~5$~km~s$^{-1}$ or; {\bf (2)} an accretion flow in which two physical 
components are present: an infalling anvelope with the central ring-like 
structure probing a compact and dense disk or torus around the central object.

Although e2-W is the only continuum source in W51e2 clearly associated with 
a UCHII region, currently e2-E seems to be the most active source in the 
region. The presence of methanol masers and the lack of a UCHII region 
point at a massive central object at an early stage of the star forming 
process.
%----------------------------------------------------------------------

%----------------------------------------------------------------------
\section*{Acknowledgements}
The work presented here is based on observations obtained with MERLIN, a 
National Facility operated by the University of Manchester at Jodrell Bank 
Observatory, on behalf of STFC.Computations were carried out using 
Legion Supercomputer, ULC. 
The authors would like to thank the referee, 
S. Ellingsen, for his valuable comments and suggestions.

%----------------------------------------------------------------------

%----------------------------------------------------------------------
\newpage
\appendix

\section{Spectra for the 6.035-GHz excited OH Zeeman components}

%%%%%%%%%%%%
%%%%%%%%%%%%
\begin{figure}
      \epsfig{file=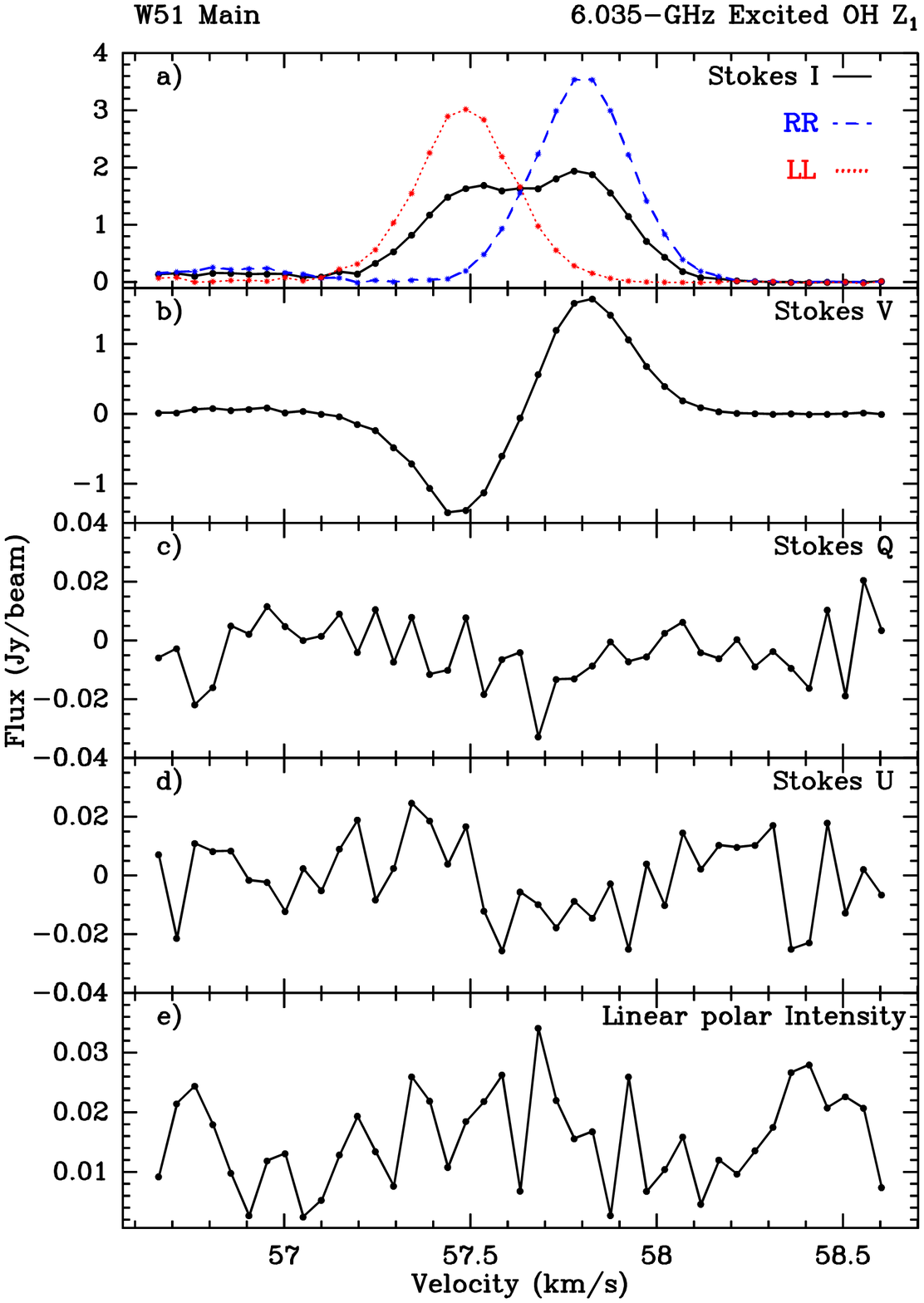,width=9.00cm}
\caption{Spectra of the Zeeman component z$_1$ (6.035-GHz excited OH line). 
         {\bf a)} Stokes~$I$, RHC and LHC spectra. {\bf b)} Stokes~$V$. 
         {\bf c)} Stokes~$Q$. {\bf d)} Stokes~$U$. 
         {\bf e)} linear polarisation intensity $P$.}
\label{fig: spectra OH 6035 z1}
\end{figure}
%%%%%%%%%%%%
%%%%%%%%%%%%

%%%%%%%%%%%%
%%%%%%%%%%%%
\begin{figure}
      \epsfig{file=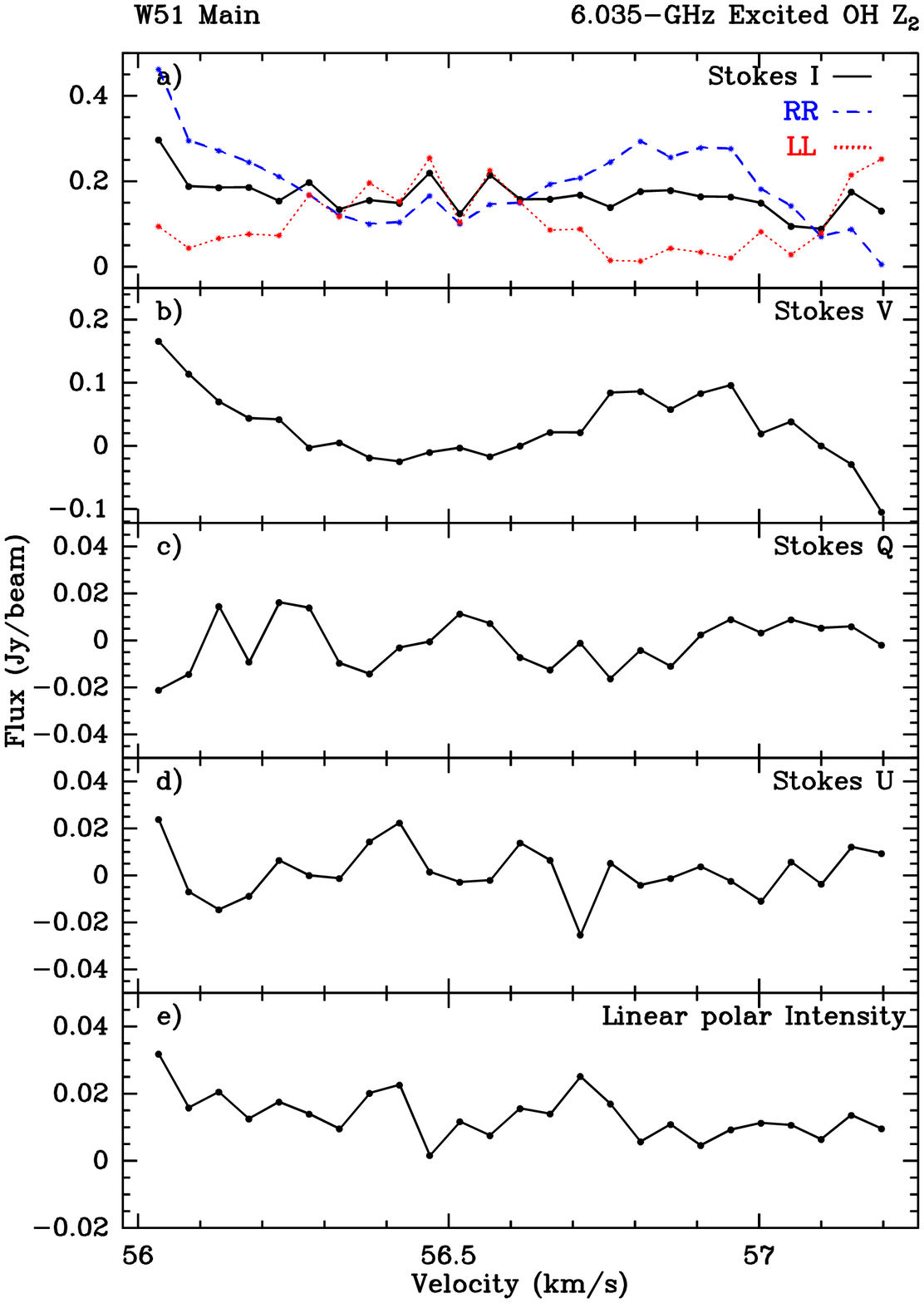,width=9.00cm}
\caption{Same as in Fig.~\ref{fig: spectra OH 6035 z1} for the Zeeman 
         component z$_2$. 
         Note that the velocity range presented here 
         encompass exclusively that of the component due to 
         contamination by other components in the adjacent channels. This is 
         the reason why the wings do not return to zero.
}
\label{fig: spectra OH 6035 z2}
\end{figure}
%%%%%%%%%%%%
%%%%%%%%%%%%

%%%%%%%%%%%%
%%%%%%%%%%%%
\begin{figure}
      \epsfig{file=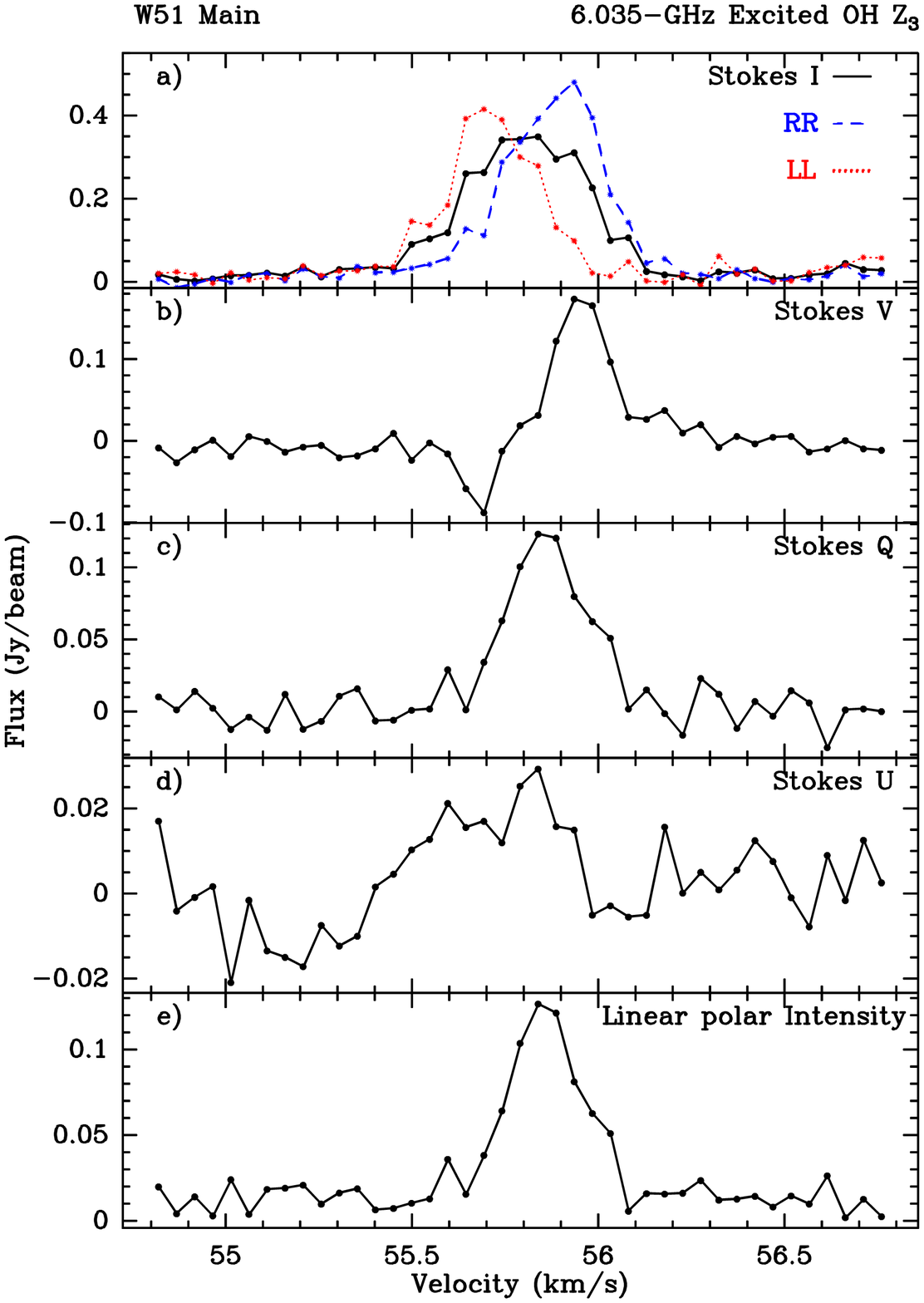,width=9.00cm}
\caption{Same as in Fig.~\ref{fig: spectra OH 6035 z1} for the Zeeman 
         component z$_3$.}
\label{fig: spectra OH 6035 z3}
\end{figure}
%%%%%%%%%%%%
%%%%%%%%%%%%

%%%%%%%%%%%%
%%%%%%%%%%%%
\begin{figure}
      \epsfig{file=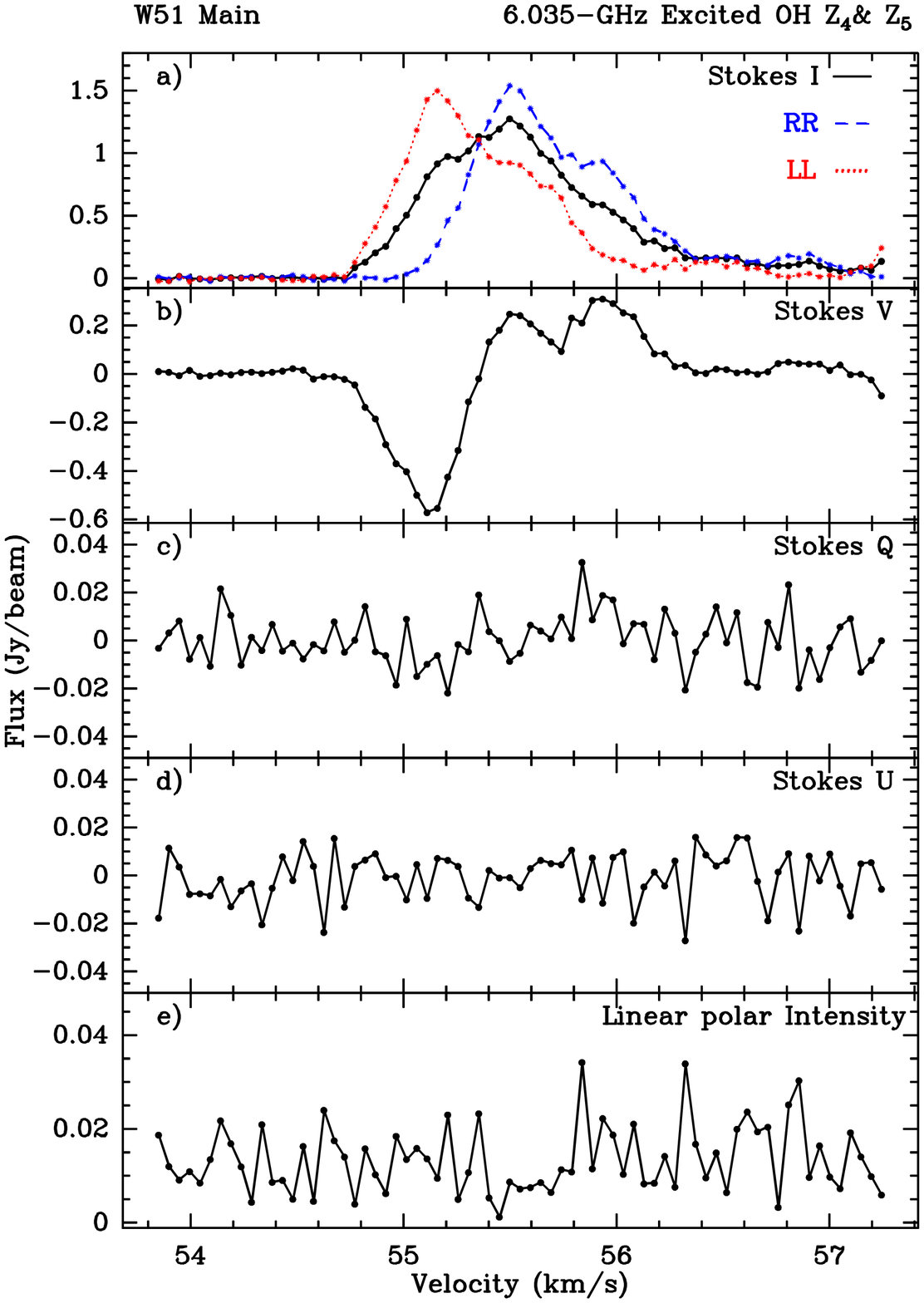,width=9.00cm}
\caption{Same as in Fig.~\ref{fig: spectra OH 6035 z1} for the Zeeman 
         components z$_4$ and z$_5$.}
\label{fig: spectra OH 6035 z4 and z5}
\end{figure}
%%%%%%%%%%%%
%%%%%%%%%%%%

%%%%%%%%%%%%
%%%%%%%%%%%%
\begin{figure}
      \epsfig{file=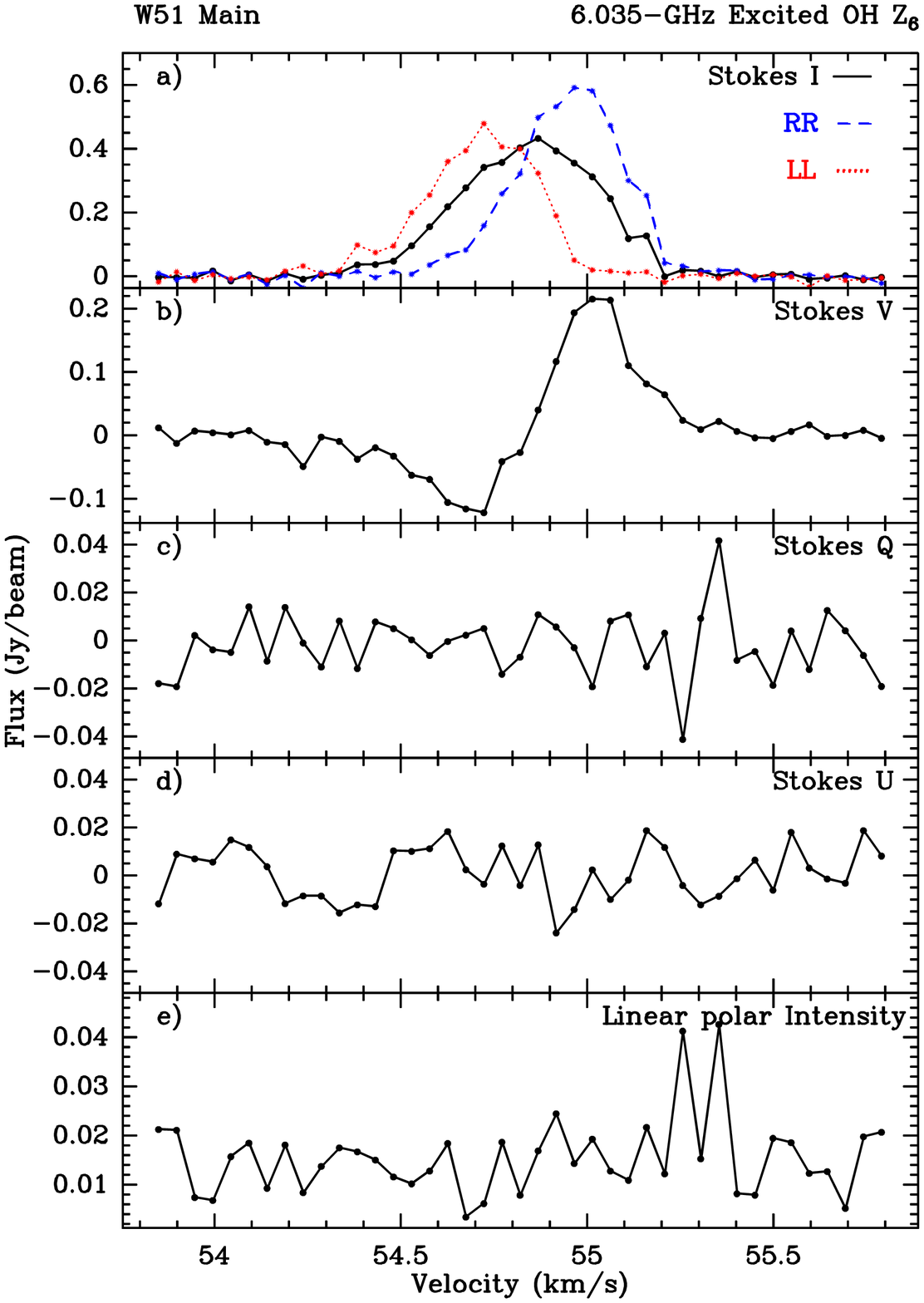,width=9.00cm}
\caption{Same as in Fig.~\ref{fig: spectra OH 6035 z1} for the Zeeman 
         component z$_6$ }
\label{fig: spectra OH 6035 z6}
\end{figure}
%%%%%%%%%%%%
%%%%%%%%%%%%

%---------------------------------------------------------------
\end{document}